\documentclass[apj,iop]{emulateapj}

\usepackage{graphicx}
\usepackage{amssymb}
\usepackage{amsmath}

\usepackage{natbib}

\def\gtorder{\mathrel{\raise.3ex\hbox{$>$}\mkern-14mu
             \lower0.6ex\hbox{$\sim$}}}
\def\ltorder{\mathrel{\raise.3ex\hbox{$<$}\mkern-14mu
             \lower0.6ex\hbox{$\sim$}}}

\usepackage{mathtools}

\slugcomment{Accepted to PASP}

\shorttitle{Survey telescopes}

\shortauthors{Ofek \& Ben-Ami}

\begin{document}

\title{Seeing-limited imaging sky surveys -- small vs. large telescopes}
\author{E.~O.~Ofek\altaffilmark{1},
S.~Ben-Ami\altaffilmark{1}
}

\altaffiltext{1}{Department of Particle Physics and Astrophysics, Weizmann Institute of Science, 76100 Rehovot, Israel.}

\begin{abstract}


Typically large telescope construction
and operation costs scale up faster than their collecting area.
This slows scientific progress, making it expensive and complicated to increase telescope size.
We review the argument that a metric that represents the capability of
an imaging survey telescopes,
and that captures a wide range of science objectives,
is the telescope {\it grasp} - the amount of volume
of space in which a standard candle is detectable per unit time.
We show that in a homogeneous Euclidean universe, and in the background-noise
dominated limit, the grasp is:
$\mathcal{G}\propto \Omega A_{\rm eff}^{3/4} \sigma^{-3/2}  t_{\rm E}^{3/4}/(t_{\rm E} + t_{D})$,
where $\Omega$ is the telescope field of view,
$A_{\rm eff}$ is the effective collecting area of the telescope,
$\sigma$ is the instrumental or atmospheric seeing
or the pixel-size, whichever dominates,
$t_{\rm E}$ is the exposure time,
and $t_{\rm D}$ is the dead time.
In this case, the optimal exposure time is three times the dead time.
We also introduce a related metric we call the {\it information-content grasp}, which summarizes the variance of all
sources observed by the telescope per unit time.
We show that, in the background-noise dominated regime, the information-content grasp scales like the grasp.
For seeing-dominated sky surveys, in terms of grasp, \'etendue, or
collecting-area optimization, 
recent technological advancements make it more
cost effective to construct multiple small telescopes rather
than a single large telescope with a similar grasp or \'etendue.
Among these key advancements are the availability of 
large-format back-side illuminated CMOS detectors with $\ltorder4\,\mu$m pixels, well suited to sample standard seeing conditions given typical focal lengths of small fast telescopes.

We also discuss the possible use of multiple small telescopes for spectroscopy and intensity interferometers.
We argue that if all the obstacles to implementing cost-effective wide-field imaging and multi-object spectrographs
using multiple small telescopes are removed, then the motivation
to build new single large-aperture ($\gtorder1$\,m) visible-light telescopes which are seeing-dominated, will be weakened.
These ideas have led to the concept of the, currently under construction, Large-Array Survey Telescope (LAST).

\end{abstract}

\keywords{
instrumentation: miscellaneous ---
methods: observational ---
telescopes}

\section{Introduction}
\label{sec:Introduction}

Some astronomical studies depend on maximizing the number of sources (of some kind)
that a telescope can observe per unit time,
or maximizing the signal-to-noise square (i.e., the variance or information content) of all sources observed per unit time.
Alternatively, we may be interested in specific objects, resolution or depth,
and in this case, either maximizing the resolution,
or/and the collecting area is required.
The first approach may be regarded as a {\it survey telescope}, and it is likely that most 
astronomical telescopes on Earth are used for this purpose
(even if they are observing one target at a time).
The second approach is applied, typically, by large collecting area telescopes --
e.g., with adaptive-optics systems, or other unique instrumentation.

Benefiting from technological advancement (e.g., large format CCDs), numerous new wide field imaging survey telescopes were recently constructed, and new machines are being built.
Examples include, 
the Palomar Transient Factory (\citealt{Law+2009_PTF}),
the Pan-STARRS survey (\citealt{Chambers+2016_PS1_Surveys}),
MASTER global robotic network (\citealt{Gorbovskoy+2013_MASTER_OpticalTelescopes}),
the All-Sky Automated Survey for Supernovae
(ASAS-SN; \citealt{Kochanek+2017_ASASSN_VarStars}),
The Gravitational wave Optical Transient Observatory
(GOTO; \citealt{Martin+2018_GOTO_Control}),
the Zwicky Transient Facility (\citealt{Bellm+2019_ZTF_Overview}),
and the Large Synoptic Survey Telescope (LSST; \citealt{Tyson+2001_LSST}).
However, there is wide room for  expanding and diversifying the global array
of survey telescopes.
One reason is that some science cases require systems that are capable
of accessing the entire sky at any given moment, while
a typical ground-based observatory can access only $\approx6$\% of the
sky, with airmass below 2, at any given time.
Among such science cases are
gravitational wave events (e.g., \citealt{Abbott+2017_GW170817_LIGO}),
fast transients (e.g., \citealt{Cenko+2013_PTF11agg}),
and exoplanets (e.g., \citealt{Charbonneau+2000_FirstTransitingExoplanet}, \citealt{Nutzman+Charbonneau2008_MEarth_DesignConsiderations}, 
\citealt{Swift+2015_Minerva_ExoPlanets_DesignCommissioning}).
These, and other science cases, call for 
around-the-globe, wide-field, high-cadence survey telescopes.
Therefore, it may be beneficial to inspect observing systems design in light of recent technological advancements.

Comparing the discovery potential of survey telescopes can guide us in the design of new survey telescopes.
A common comparison metric used in the literature
is the \'etendue, which is proportional to the 
amount of flux a single telescope receives from all sources.
The \'etendue is given by $A\Omega$,
where $A$ is the collecting area
and $\Omega$ is the field of view (e.g., \citealt{Tyson+2001_LSST}).
\cite{Tonry2011_ATLAS_SurveyCapability} 
suggested another metric which he called the survey capability and is proportional to $\sum_{i}{(S/N)_{i}^2}$, where $(S/N)_{i}$ is the sources signal-to noise.
This definition, with some minor differences, is similar to the information-content grasp we introduce in \S\ref{sec:Info}.
Here we derive the information-content grasp analytically as a function of the telescope parameters, and we also show that the information content grasp is equivalent to the grasp.”

\cite{Djorgovski2012_SkySurveysReview} suggested a figure of merit which is based on the inverse of the telescope limiting flux (see Eq.~\ref{eq:flim}).
\cite{Pepper+2002_SmallTelescopeTransits} and \cite{Bellm2016_VolumetricRate} advocated the use of a volumetric rate metric
(for which we adopt the name {\it grasp}),
which is equivalent to counting the number of sources,
with a constant space density in an Euclidean universe, that a telescope can observe per unit time.

\cite{Pepper+2002_SmallTelescopeTransits}
analyzed the following problem: Given a single telescope
with a fixed focal ratio and detector size,
what is the optimal aperture size for conducting a survey for
hot-Jupiter transiting planets.
They concluded that in order to maximize the number of detected
hot-Jupiter transiting planets around bright stars, it is best to use a $1$-inch diameter telescope.
This paper likely inspired the design and construction of some
successful surveys based on small telescopes aimed at detecting and characterizing transiting planets
(e.g., HAT, \citealt{Bakos+2004_HAT}; WASP, \citealt{Pollacco+2006_WASP_superWASP}).
In fact, the approach of multiple small telescopes for sky surveys
is becoming very popular.
Such telescopes are used for multiple objectives,
including transiting planets (e.g., EvryScope, \citealt{Law+2015_Evryscope_ScienceCase}; Next Generation Transit Survey [NGTS], \citealt{Wheatley+2018_NGTS});
transients (e.g., ASAS-SN, \citealt{Kochanek+2017_ASASSN_VarStars}; 
GOTO, \citealt{Martin+2018_GOTO_Control};
ATLAS, \citealt{Heinze+2018_ATLAS_VarStars};
MASTER, \citealt{Gorbovskoy+2013_MASTER_OpticalTelescopes}),
low-surface brightness galaxies (DragonFly, \citealt{Abraham+2014_DragonFly}; \citealt{Danieli+2018_DwarfGalaxiesInTheField_DragonFly}),
and asteroids (e.g., \citealt{Heinze+2018_ATLAS_VarStars}).

\cite{Bellm2016_VolumetricRate} clearly identified the volumetric rate as useful for
comparing survey telescopes, and calculated it from the limiting magnitude for various sky surveys.
Furthermore, \cite{Bellm+2019_ZTF_Scheduler} used the volumetric rate
to optimize the Zwicky Transient Facility observing schedule.

It is likely too simplistic to describe different science cases with
a single metric like the grasp.
For example, sometimes we care about cadence,
and not only the volume of space probed.
This, however, can be addressed by assuming that we are running out of sky
after some fixed amount of time (see e.g., \S\ref{sec:outsky}).
Other complications include the geometry of the universe,
and the nonconstant space density of sources (see e.g., \S\ref{sec:universe}).
In some cases, we may be interested in the information content (i.e., variance) of all the observable sources --
for that reason, we also introduce the {\it information-content grasp}
(see \S\ref{sec:Info}).
Such details should be taken into account when comparing survey performances.
Nevertheless, we argue that the grasp is an informative metric,
as it gives us a rough scaling of the number of sources
a survey telescope can detect per unit time.

Here we derive the volumetric-rate (grasp or survey speed) formula and the information-content-grasp formula.
We derive an analytic expression for the optimal integration time,
given the dead time,
and we extend the grasp formula to diffraction-limited telescopes.
We explore several variants of these scaling relations,
including the limit where the survey is running out of sky
and the effects of a non-Euclidean universe.
We mainly deal with the simplest form of the grasp
and how it scales with the various system parameters.
%
We define the cost-effectiveness of a survey telescope as its grasp per unit construction cost (and alternatively, maintenance cost).
We argue that due to the recent availability of large format CMOS detectors with small pixels well suited to sample standard seeing conditions at focal lengths of $\mathcal{O}(10^3)$\,mm. it is possible that (arrays of) small telescopes are becoming considerably (about an order of magnitude) more cost effective compared to most existing and planned wide-field imaging telescopes.

In \S\ref{sec:grasp}, we derive the grasp formula analytically,
along with several variants including the information-content grasp.
In \S\ref{sec:cost}, we define the survey cost-effectiveness,
with practical considerations discussed in \S\ref{sec:practice}.
In \S\ref{sec:spec}, we briefly discuss the possibility of using
multiple small telescopes for spectroscopy and interferometry,
and we conclude in \S\ref{sec:conc}.

\section{The grasp}
\label{sec:grasp}

The {\it grasp} of a telescope system
is proportional to the volume the telescope can probe per unit time.
The actual volume depends on the intrinsic luminosity of the sources,
but if we care about comparing two optical systems, or analyzing
the grasp scaling with various parameters, the intrinsic luminosity or the shape
of the luminosity function becomes irrelevant\footnote{As long as the power-law index of the luminosity function
is the same over the relevant luminosity range.}.

The grasp formula can be derived by inspecting the signal-to-noise ($S/N$) ratio formula.
For source detection or, alternatively, at the limit of a background-dominated noise,
the $S/N$ of a circularly symmetric Gaussian point spread function (PSF) is given by\footnote{The $(S/N)^{2}$ is an additive quantity (e.g., \citealt{Zackay+2017_CoadditionI}). Therefore, we simply integrate over the $(S/N)^{2}$ of each pixel in the PSF.}:
\begin{align}
    & \Big(\frac{S}{N}\Big) = \sqrt{\int_{0}^{\infty}{2\pi r dr \frac{F^{2}A_{\rm eff}^{2}t_{\rm E}^{2} }{B A_{\rm eff} t_{\rm E} }\frac{e^{-r^2/(\sigma^{2})} }{4\pi^{2}\sigma^{4}}  }}  \nonumber \\
    & = \frac{F A_{\rm eff} t_{\rm E}}{\sqrt{4\pi\sigma^{2} B A_{\rm eff} t_{\rm E} }}.
    \label{eq:SN}
\end{align}
Here, $S/N$ is the signal-to-noise ratio required for source detection
or for some measurement in the background-noise dominated regime,
$F$ is the flux of the source in units of number of photons per unit time per unit area,
$A_{\rm eff}$ is the aperture effective\footnote{The effective area is proportional to the actual collecting area, filter width and system throughput.} collecting area,
$t_{\rm E}$ is the integration time,
$B$ is the sky background in units of number of photons, per unit area, per unit time,
per unit solid angle of the sky,
and $\sigma$ is the width of the Gaussian PSF (e.g., imposed by the atmospheric seeing,
optical abberations, the detector, or the angular size of the object of interest).
%
We note that for a measurement process (rather than detection)
in the source-noise-dominated limit (e.g., exoplanet searches), or the read-noise
dominated limit, Equation~\ref{eq:SN} is no longer correct
(see general formula and derivation in \citealt{Zackay+2017_CoadditionI}).
The source-noise dominated case is discussed in \S\ref{sec:sourcenoise}.

Next, we can isolate $F$ to find the limiting flux of the system:
\begin{equation}
    F_{\rm lim} = \sqrt{4\pi}\Big(\frac{S}{N}\Big) B^{1/2} \sigma (A_{\rm eff} t_{\rm E})^{-1/2}.
    \label{eq:flim}
\end{equation}
Since $d\propto F^{-1/2}$,
where $d$ is the distance to the source,
the distance to which we can detect a standard candle is
\begin{equation}
    d \propto \Big(\frac{S}{N}\Big)^{-1/2} B^{-1/4} \sigma^{-1/2} (A_{\rm eff} t_{\rm E})^{1/4},
\end{equation}
and the volume in which we can detect sources is
\begin{equation}
    V \propto \Omega \Big(\frac{S}{N}\Big)^{-3/2} B^{-3/4} \sigma^{-3/2} (A_{\rm eff} t_{\rm E})^{3/4}.
\end{equation}
Here, $\Omega$ is the field of view of the telescope.

Next, we are interested in the volume per unit time (i.e., the grasp; $\mathcal{G}=V/(t_{\rm E} + t_{\rm D})$), where $t_{\rm D}$ is the dead time
(e.g., slew time, readout time), which is given by
\begin{equation}
    \mathcal{G} \propto \Omega \Big(\frac{S}{N}\Big)^{-3/2} A_{\rm eff}^{3/4} B^{-3/4} \sigma^{-3/2} \frac{t_{\rm E}^{3/4}}{t_{\rm E} + t_{\rm D}  }.
    \label{eq:grasp_basic}
\end{equation}
Hereafter, we will refer to this as the Grasp Equation.

It is important to note that the Grasp Equation
is valid under the assumptions mentioned earlier.
For example, for a system of small telescope with small enough
dead time (and hence short optimal exposure), 
the telescope may not be able to detect even the brightest target of some class (e.g., supernova).
Since telescopes which are large enough to be seeing dominated,
have aperture area ranging over 2--3 orders of magnitude,
in order to compare the smallest telescope to the largest telescope,
our assumptions about the continuity of the luminosity function,
source homogeneity, and Euclidean space must be valid over the appropriate range.
In some of the next subsections, we consider more complicated cases.


\subsection{The optimal ratio between the exposure time and the dead time}

In the case that we are interested in observing the sky at some cadence, and assuming
we are not running out of sky (e.g., \S\ref{sec:outsky}) and given a dead time $t_{\rm D}$, the Grasp Equation has a maximum
in respect to the exposure time $t_{\rm E}$.

Given Equation~\ref{eq:grasp_basic}, with an arbitrary power-law of $(A_{\rm eff}t_{\rm E})^{\alpha}$, we can calculate the optimal exposure time (to maximize the grasp)
in units of the dead time. Since
\begin{equation}
    \frac{d}{dt_{\rm E}}\Big(\frac{t_{\rm E}^{\alpha}}{t_{\rm E} + t_{\rm D}  }  \Big) = \frac{t_{\rm E}^{\alpha-1}(\alpha t_{\rm D} - t_{\rm E}[1-\alpha])}{(t_{\rm D} + t_{\rm E})^{2}},
\end{equation}
the optimum integration time that maximizes the observed volume per unit time is (for $1>\alpha>0$)
\begin{equation}
    t_{\rm E, optimal} = \frac{\alpha t_{\rm D}}{1-\alpha},
\end{equation}
and in the special case of $\alpha=3/4$, we get
\begin{equation}
    t_{\rm E, optimal} = 3t_{\rm D}.
\end{equation}
For $\alpha$ approaching 1, $t_{\rm E, optimal}/t_{\rm D}$ approach infinity, see section $ 2.6$ discussing Non-Euclidean geometries.

\subsection{The diffraction-limited case}

In the diffraction-limited case, where the collecting aperture is circularly symmetric
(see, however, \citealt{Nir+2019_NoncircularPupilTelescope}), $\sigma\propto A^{-1/2}$
and we get
\begin{equation}
    \mathcal{G} \propto \Omega A_{\rm eff}^{3/2} B^{-3/4} \frac{t_{\rm E}^{3/4}}{t_{\rm E} + t_{\rm D}  }.
\end{equation}
Therefore, a diffraction-limited sky survey grasp grows considerably faster than $A$.
The reason is that while the background contribution to the source increases with $A$, the background area
contributing to the source decreases with $A$.
We note that not every telescope that provides diffraction-limited imaging will have a grasp scaling like
$A_{\rm eff}^{3/2}$.
For example, speckle-interferometry-like methods (\citealt{Labeyrie1970_SpeckleInterferometry}; \citealt{Bates1982_SpeckleInterferometryReview}; \citealt{Zackay+2017_CoadditionII}) effectively combine the light from multiple
speckles,
but also combine the background from these speckles.




\subsection{Running out of sky}
\label{sec:outsky}

There are two ways to increase the volume probed by multiple ($N$) small telescopes.
They can either observe the same field of view, in which case the grasp
will increase like $N^{3/4}$,
or they can observe different fields, in which case the grasp will increase like $N$.

If we choose to maximize the grasp, we will need to cover more sky area.
However, at some point, a system of small telescopes will run out of sky,
and in this case, the efficiency will transition from being proportional to $N$ to being proportional to $N^{3/4}$.

For a system in which $\Omega_{1}$ is the field of view of a single telescope,
$N$ is the number of telescopes, and we observe
different fields until we run out of sky\footnote{In a similar manner, we can introduce running out of time -- e.g., if we are interesting in observing the sky in a specific cadence.} ($\Omega_{\rm sky}$), we can write:
\begin{equation}
    \mathcal{G} \propto \Big(\frac{A_{\rm eff}}{B\sigma^{2}}\Big)^{3/4} \frac{t_{\rm E}^{3/4}}{t_{\rm E} + t_{\rm D}  } \times
    \begin{cases*}
         N \Omega_{1}    &{\rm if}~$N\Omega_{1} < \Omega_{\rm sky}$ \\
         \Omega_{\rm sky} \Big(\frac{N\Omega_{1}}{\Omega_{\rm sky} }  \Big)^{3/4}      &{\rm otherwise}.
    \end{cases*}
\end{equation}
For some systems (e.g., EvryScope; \citealt{Law+2015_Evryscope_ScienceCase}; \citealt{Ratzloff+2019_EvryScope}) the transition will happen very quickly.

\subsection{Source density function}

So far, we assumed the sources have a uniform density with distance.
Next, we assume that the source density has some power-law distribution with distance:
\begin{equation}
    \rho(d) = \rho_{0}(d/d_{0})^{\beta},
\end{equation}
where $\rho_{0}$ and $d_{0}$ are the density
and distance normalization, and $\beta$ is
the power-law index of the density distribution.

Integrating over $d$, for $\beta>-3$ (for a smaller $\beta$, the density diverges), the grasp is given by
\begin{equation}
    \mathcal{G} \propto \Omega \frac{1}{3+\beta} \Big(\frac{S}{N}\Big)^{-(3+\beta)/2} (B^{-1} A_{\rm eff} t_{\rm E})^{(3+\beta)/4} \frac{1}{t_{\rm E} + t_{\rm D}}.
\end{equation}
We see that, in an Euclidean universe, for $\beta>1$, the power-law dependence on $A_{\rm eff}$ becomes steeper than the power-law dependence on $\Omega$ (i.e., $+1$).

\subsection{Non-Euclidean universe}
\label{sec:universe}

\begin{figure}
    \centering
    \includegraphics[width=7cm]{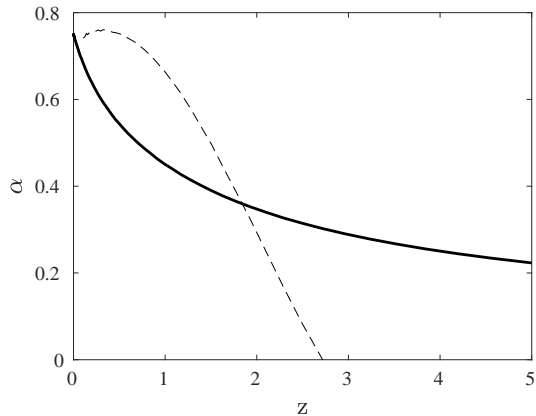}
    \caption{$\alpha$ (the power-law index of $A_{\rm eff}t_{\rm E}$) as a function of redshift (solid line), assuming {\it Planck} cosmological parameters (\citealt{Ade+2016_Planck2015_CosmologicalParameters}). The solid line represents $\alpha$ in the case of sources with a uniform number per comoving volume, while the dashed line is the approximate $\alpha$ when we introduce the measured evolution in the star formation rate (\citealt{Cucciati+2012_SFRz}). This plot ignores the evolution in the luminosity function, redshift of the spectrum, and time dilation. For example, the time dilation may be important for transient sources.}
    \label{fig:alpha_z}
\end{figure}

All the scaling relations we derived so far assume a Universe with an Euclidean geometry and no time-dilation.
However, these assumptions break down quickly when the redshift $z$ of the sources we are interested in is large.
In addition, some cosmological sources have a density function with a steep dependency on redshift (e.g., the star formation rate;
\citealt{Cucciati+2012_SFRz}).
To parameterize the effect of cosmology,
we approximate the power-law dependence of the grasp on the collecting area using a power-law with index $\alpha$ -- i.e,
\begin{equation}
    \mathcal{G} \propto \Omega A_{\rm eff}^{\alpha} \frac{t_{\rm E}^{\alpha}}{t_{\rm E} + t_{\rm D}  },
\end{equation}
where $\alpha$ is the power-law index of $(A_{\rm eff} t_{\rm E})$.
In Figure~\ref{fig:alpha_z} we show a numerical estimation of $\alpha$ as a function of redshift.
The solid line shows $\alpha$ assuming a constant source density (and no luminosity evolution),
while the dashed line assumes the source density follows the star formation rate (e.g., \citealt{Cucciati+2012_SFRz})
but disregards any evolution in the luminosity function.
We stress that the power-law representation over a wide range of telescope sizes,
is only a rough approximation.
In addition, for some redshift ranges $\alpha<0$ and our approximation breaks.
We have to be careful not to over interpret this plot.
In reality, there are several additional complications.
For example, for transient sources (e.g., supernovae), the rate of events
further decreases like $1/(1+z)$, due to the time dilation of the transient rate.
Moreover, this plot ignores the fact that the luminosity function of sources may change with redshift.

\subsection{The source-noise dominated case}
\label{sec:sourcenoise}

So far we have discussed the background-noise dominated regime.
For bright sources (\textit{e.g.,} exoplanet transit detection around FGK stars), the source noise is the dominating noise term. The criterion for this case\footnote{For cases in which the readnoise is sub dominant.} is
given by comparing the numerator and the square of the denominator
in Equation~\ref{eq:SN}:
\begin{equation}
    F_{\rm t}\gtorder 4\pi B \sigma^{2},
    \label{eq:F_t}
\end{equation}
where $F_{\rm t}$ is the flux above which the transition from background-noise to source-noise occurs.
For example, for $\sigma\approx0.28''$, $1''$, $2''$, $10''$, the transition
between background-noise dominated and source-noise dominated
occurs at about 0, 2.7, 4.2, and 7.7 magnitudes below
the sky surface magnitude (in units of magnitude per square arcseconds), respectively.

In the source-noise-dominated case:
\begin{equation}
    \Big(\frac{S}{N}\Big) = \sqrt{FA_{\rm eff}t_{\rm E}}.
    \label{eq:sn_sourcenoise}
\end{equation}

Repeating the steps in \S\ref{sec:grasp} with Equation~\ref{eq:sn_sourcenoise} we find
\begin{equation}
    \mathcal{G} \propto \Omega \Big(\frac{S}{N}\Big)^{-3} A_{\rm eff}^{3/2} t_{\rm E}^{3/2} \frac{1}{t_{\rm E} + t_{\rm D}}.
    \label{eq:grasp_sourcenoise}
\end{equation}
This suggests that deep in the source-noise-dominated case, and for constant $S/N$, larger telescopes are preferable,
while the seeing is not important.

\subsection{The information-content grasp}
\label{sec:Info}

For some applications we care about measuring some properties
like the flux (e.g., exoplanets search),
first moment (e.g., astrometry), or shape (e.g., weak lensing),
as precisely as possible.
Assuming that we are limited by Poisson noise,
we can define the {\it information-content grasp},
as the sum of variances\footnote{The Fisher information of a random variable is proportional to the variance of this random variable.}
of all sources in the field of view.

For sources with a flux density function of
the form
\begin{equation}
    \frac{dN}{dF}\propto F^{-\zeta},
\end{equation}
where $N$ is the number of sources, $F$ is the flux,
and $\zeta$ is a power-law index,
the volumetric information content integrated
between the flux limit ($F_{\rm lim}$)
and some upper-flux bound ($F_{\rm max}$; e.g.,
saturation limit) is
\begin{equation}
    V_{\rm I} \propto \Omega\int_{F_{\rm lim}}^{F_{\rm max}}{\Big(\frac{S}{N} \Big)_{\rm m}^{2} F^{-\zeta} dF}.
    \label{eq:VI}
\end{equation}
Here $(S/N)_{\rm m}^{2}$ is the variance of a measurement process
(see general formula in \citealt{Zackay+2017_CoadditionI}).
For simplicity we will treat only the background dominated case,
in which
\begin{equation}
    (S/N)_{\rm m}^{2} = \frac{F^{2}}{4\pi\sigma^{2}B} A_{\rm eff}t_{\rm E}.
    \label{eq:SN2_back}
\end{equation}
Note that in the background-noise dominated case,
the $S/N$ for detection and measurement are the same.
Plugging Equation~\ref{eq:SN2_back} into Equation~\ref{eq:VI},
setting $x\equiv F_{\rm max}/F_{\rm lim}$,
and integrating, we get (for $\zeta\ne 3$) the information-content volume
\begin{equation}
    V_{\rm I} = \Omega \frac{(x^{3-\zeta}-1)}{4\pi(\zeta - 3)}
    \frac{A_{\rm eff} t_{\rm E}}{\sigma^{2}B} F_{\rm lim}^{3-\zeta}.
    \label{eq:infocontent_V}
\end{equation}
Here $x$ can be regarded as the dynamic range of the observations,
and we assume that we are in the background-noise dominated regime in
the entire dynamic range.
To get the information-content grasp , we plug Equation~\ref{eq:flim} into Equation~\ref{eq:infocontent_V} and divide by the exposure time plus the dead time.
\begin{equation}
    \mathcal{G}_{\rm I} \propto \Omega \frac{(x^{3-\zeta}-1)}{4\pi(\zeta - 3)}
    \Big(\frac{S}{N} \Big)^{3-\zeta} \sigma^{1-\zeta} 
    \Big(\frac{A_{\rm eff}}{B}\Big)^{(\zeta-1)/2}
    \frac{t_{\rm E}^{(\zeta-1)/2}}{t_{\rm E}+t_{\rm D}}.
    \label{eq:infocontent_grasp}
\end{equation}

In the case of an Euclidean Universe with a homogeneous source density ($\zeta=5/2$),
and in the background-noise dominated limit,
the information-content grasp grows like $\sigma^{-3/2}A_{\rm eff}^{3/4}$.
I.e., the information-content grasp scales like the  grasp.

\section{Telescope cost-effectiveness}
\label{sec:cost}

We define the cost-effectiveness of a survey telescope as the grasp per unit cost.
According to \cite{vanBelle+2004_TelescopeCostScaling}, the cost of telescopes grows like $\sim A^{\gamma}$, with $\gamma\approx1.15$, but the exact power-law ranges from
$\gamma\approx1$ to $1.5$, and depends on the telescope details.
Furthermore, with newer telescope design, segmented mirrors and shorter focal-ratio
implemented in new large telescopes, it is possible that $\gamma$ for large telescopes is slightly smaller than 1.
Adopting the basic grasp formula,
or even the \'etendue, suggests that
for seeing-dominated survey telescopes,
multiple small telescopes
have the potential to be more cost-effective than one single large telescope
with the same grasp (i.e., $N\Omega_{S}A_{S}^{3/4} = \Omega_{L}A_{L}^{3/4}$;
subscripts $S$ and $L$ correspond to small and large telescopes, respectively).
For such equivalent
systems\footnote{One possible difference between large and small telescopes is that the seeing of large telescope maybe better by about 10\% (e.g., \citealp{Martinze+2010_LargeTelescopeSeeing}).},
the cost-effectiveness of a small telescope system,
compared to a large telescope with the same grasp,
is proportional to $\sim N(\Omega_{S}/\Omega_{L}) (A_{S}/A_{L})^{3/4-\gamma}$.
This suggest a factor of a few at most between the cost effectivness of small and large telescopes.
It is possible, however, that a larger factor in cost effectivness of small vs. large telescopes hides in mass production, and the use of products which are not unique for professional astronomy.


Given the fast development and progress in detector technology,
optics manufacturing technology and computing, as well as the large-scale market economy,
when designing a new system, it is worthwhile to compare the capabilities of large vs. small telescopes.
This comparison depends on the details of our science goals.

Our rough analysis suggests that the cost-effectiveness of some small telescope systems
(e.g., GOTO, \citealt{Martin+2018_GOTO_Control}; ATLAS, \citealt{Tonry2011_ATLAS_SurveyCapability}, \citealt{Heinze+2018_ATLAS_VarStars})
is probably\footnote{Based on information available on the Internet.} 
higher by a factor of about 3 compared to some of the other small or big survey telescopes.
An important question is: Can we further improve the cost effectiveness of telescopes?
We argue that technological development of the past several years have likely made it possible to increase
the grasp-based cost-effectiveness of seeing-limited imaging survey telescopes by about an order of magnitude.
One such development
is the fast progress made in the quality of CMOS detectors,
which are considerably less expensive than CCD-based cameras,
and are available with small pixels (see \S\ref{sec:practice}).



\section{Practical considerations}
\label{sec:practice}

In order to maximize the cost-effectiveness of seeing-limited survey
telescopes, given the strong dependence
of the grasp on the image quality, it is desirable that the system will be limited
by the seeing\footnote{One possible exception is the case in which we are interested in extended sources (e.g., \citealt{Abraham+2014_DragonFly}).}.
For a 20\,cm telescope, the visible-light diffraction limit is
about $0.7''$, so our first requirement is for telescopes with
a diameter $D\gtorder20$\,cm.
Next, we require optics that provide an image quality
comparable to the seeing and a pixel size that critically
samples the PSF\footnote{The Nyquist frequency is not well defined for a Gaussian PSF. However, for practical purposes, we suggest using a pixel scale of about 2.3 pixels across the PSF full width at half maximum (see e.g., \citealt{Ofek2019_Astrometry_Code}).}.
%

Since we would like to maximize the field of view,
systems with a short focal length are required.
This means that in order to critically sample
the PSF, we need small pixels.
For example, with a 30\,cm f/2 telescope,
$3$\,$\mu$m pixels are required in order to have a pixel
scale of $1''$/pix.
Most importantly, such large-format
back-side illuminated CMOS detectors
with small pixels
have only now become available.
Another important point is that CMOS devices are considerably less expensive
than equivalent CCDs, and that CMOS technology is maturing fast.
Therefore, utilizing CMOS technology has the potential to further increase the cost-effectiveness of survey telescopes.
The pixel-size requirement means that fully utilizing
the cost-effectiveness of small telescopes
has only recently become possible.
Another important factor is that
small telescopes and cameras have a huge market,
and this has an important impact on their availability,
and cost-effectiveness.

The multiple-small-telescopes-for-sky-survey approach has
several challenges.
First,
operating and analyzing data from a large number of telescopes
requires considerable computing power and data storage.
Saving all the raw data from a large ($\mathcal{O}[1000]$)
number of telescopes is challenging.
However, a possible solution is to process
the data and save selected data products.
Second, multiplicity usually comes with some complexity
(e.g., maintenance).
Therefore, we require a proof that the small telescope approach
has the potential to increase the cost-effectiveness, of existing
and planned sky-survey systems, by an order of magnitude.
%
%
%

\section{Comments about spectroscopy and interferometry}
\label{sec:spec}

So far we have focused on imaging surveys.
It is of interest to consider extending the multiple-small telescope approach
to spectroscopy and maybe intensity interferometers
(\citealt{Brown+Twiss1956_HBT_CorrelationsBetweenPhotons,Brown+Twiss1957_HBT_Interferometer,Brown+Twiss1958_HBT_Interferometer_Astronomy}).
Applying this approach to spectroscopy was already discussed in the literature (e.g., \citealt{Eikenberry+2019_PolyOculus}).

The grasp formula represents a special case in which we are
interested in all, or a subset, of objects in a volume
of space.
However, in spectroscopy we are typically limited to a fixed number of targets per exposure. Therefore, when discussing spectroscopy, we ignore the gras formula.
Nevertheless, it is interesting to understand if constructing
a telescope for spectroscopy
may be more cost-effective using multiple small telescopes
compared to a single telescope with the same collecting area.

Spectroscopic observations using multiple small telescopes
requires either a spectrograph per telescope,
or combining the light from many telescopes into
a single spectrograph (e.g., \citealt{Eikenberry+2019_PolyOculus}).
Furthermore, one may consider a single object spectrograph,
or a multi-object spectrograph.
For the spectrograph-per-telescope option,
since a single efficient spectrograph may be at least an order of
magnitude more expensive than a 20\,cm-class telescope,
the optimum cost-effective system
may require a larger telescope (e.g., 0.5--1\,m class).
The second option, of efficiently combining the light
from multiple small telescopes into a single
spectrograph, is challenging.
Among the possibilities to combine the light from multiple
telescopes is using a variant of the optical multiplexer (\citealt{Zackay+2014_MultiplexerAlgo}, \citealt{Ben-Ami+2014_MultiplexerInstrument}).
Another light-combibibg method was suggested for the PolyOculus project (\citealt{Eikenberry+2019_PolyOculus}).

\section{Summary}
\label{sec:conc}

We define the {\it grasp} as the relative volume of space that can be observed by a telescope system
per unit time (i.e., survey speed).
We argue that the {\it grasp} of a seeing-limited imaging-survey telescope,
even in its simplest form, is a good representation of
the scientific return from the system. We derive the grasp formula, and discuss it mainly in the limit
of seeing-dominated surveys assuming an Euclidean universe with a uniform source density.
We also discuss a related quantity we call the
{\it information-content grasp}
that measures the variance of all sources in
the field of view, per unit time.

We argue that, given the power-law dependence of the grasp on the telescope collecting area,
combined with recent technological developments,
it is now likely more cost-effective to build multiple small telescopes than a single large telescope,
with the same grasp and even with the same collecting area.
Specifically, the availability of high-quality backside illuminated CMOS devices with small pixels, a large format,
and a fast readout may be a game-changer in the near future.
Furthermore, these CMOS devices are typically a factor of
a few less expensive per unit area,
have simpler electronics, and consume less power,
compared with the CCD technology.
Adopting this approach in astronomy, has the potential to increase the cost-effectiveness of
seeing-dominated survey telescopes by an order of magnitude, which
eventually  may enable faster astronomical research progress.
We argue that if all the obstacles to implementing cost-effective wide-field imaging and multi-object spectrographs
using multiple small telescopes are removed, then the motivation
to build new single large-aperture ($\gtorder1$\,m) visible-light telescopes which are seeing-dominated, will be weakened.

In order to test this ideas
we are constructing the
the Large-Array Survey Telescope (LAST) project
(Ben-Ami et al. in prep.).
A LAST node will be constructed from 48, 27-cm f/2.2 telescopes.
Each telescope, equipped with a back-side illuminated CMOS device,
will have a field of view of about 7.4\,deg$^{2}$ and pixel scale of $1.2''$.
So a LAST node is equivalent to a 27\,cm telescope with field-of-view of about 355\,deg$^{2}$,
or a 1.9\,m telescope with field-of-view of 7.4\,deg$^{2}$.
The goal of the LAST project is
not only to conduct scientific investigation, but also
to provide a proof-of-concept that it is possible
to increase the cost-effectiveness of survey telescopes by an order of magnitude,
compared to existing and under-construction systems.

\acknowledgments

We thank Shri Kulkarni, Jason Spyromilio, Avishay Gal-Yam, Barak Zackay,
and Guy Nir for valuable discussions.
E.O.O. is grateful for support by
grants from the 
Willner Family Leadership Institute, Madame Olga Klein - Astrachan,
André Deloro Institute,
Schwartz/Reisman Collaborative Science Program,
Paul and Tina Gardner,
The Norman E Alexander Family Foundation ULTRASAT Data Center Fund,
Jonathan Beare,
Israel Science Foundation,
Minerva, BSF, BSF-transformative,
and the Weizmann-UK.
S.B.A is grateful for the support of the Azrieli Foundation.

\bibliography{papers.bib}

\begin{thebibliography}{41}%
\makeatletter
\providecommand \@ifxundefined [1]{%
 \@ifx{#1\undefined}
}%
\providecommand \@ifnum [1]{%
 \ifnum #1\expandafter \@firstoftwo
 \else \expandafter \@secondoftwo
 \fi
}%
\providecommand \@ifx [1]{%
 \ifx #1\expandafter \@firstoftwo
 \else \expandafter \@secondoftwo
 \fi
}%
\providecommand \natexlab [1]{#1}%
\providecommand \enquote  [1]{``#1''}%
\providecommand \bibnamefont  [1]{#1}%
\providecommand \bibfnamefont [1]{#1}%
\providecommand \citenamefont [1]{#1}%
\providecommand \href@noop [0]{\@secondoftwo}%
\providecommand \href [0]{\begingroup \@sanitize@url \@href}%
\providecommand \@href[1]{\@@startlink{#1}\@@href}%
\providecommand \@@href[1]{\endgroup#1\@@endlink}%
\providecommand \@sanitize@url [0]{\catcode `\\12\catcode `\$12\catcode
  `\&12\catcode `\#12\catcode `\^12\catcode `\_12\catcode `\%12\relax}%
\providecommand \@@startlink[1]{}%
\providecommand \@@endlink[0]{}%
\providecommand \url  [0]{\begingroup\@sanitize@url \@url }%
\providecommand \@url [1]{\endgroup\@href {#1}{\urlprefix }}%
\providecommand \urlprefix  [0]{URL }%
\providecommand \Eprint [0]{\href }%
\providecommand \doibase [0]{http://dx.doi.org/}%
\providecommand \selectlanguage [0]{\@gobble}%
\providecommand \bibinfo  [0]{\@secondoftwo}%
\providecommand \bibfield  [0]{\@secondoftwo}%
\providecommand \translation [1]{[#1]}%
\providecommand \BibitemOpen [0]{}%
\providecommand \bibitemStop [0]{}%
\providecommand \bibitemNoStop [0]{.\EOS\space}%
\providecommand \EOS [0]{\spacefactor3000\relax}%
\providecommand \BibitemShut  [1]{\csname bibitem#1\endcsname}%
\let\auto@bib@innerbib\@empty
\bibitem [{\citenamefont {{Law}}\ \emph {et~al.}(2009)\citenamefont {{Law}},
  \citenamefont {{Kulkarni}}, \citenamefont {{Dekany}}, \citenamefont {{Ofek}},
  \citenamefont {{Quimby}}, \citenamefont {{Nugent}}, \citenamefont {{Surace}},
  \citenamefont {{Grillmair}}, \citenamefont {{Bloom}}, \citenamefont
  {{Kasliwal}}, \citenamefont {{Bildsten}}, \citenamefont {{Brown}},
  \citenamefont {{Cenko}}, \citenamefont {{Ciardi}}, \citenamefont {{Croner}},
  \citenamefont {{Djorgovski}}, \citenamefont {{van Eyken}}, \citenamefont
  {{Filippenko}}, \citenamefont {{Fox}}, \citenamefont {{Gal-Yam}},
  \citenamefont {{Hale}}, \citenamefont {{Hamam}}, \citenamefont {{Helou}},
  \citenamefont {{Henning}}, \citenamefont {{Howell}}, \citenamefont
  {{Jacobsen}}, \citenamefont {{Laher}}, \citenamefont {{Mattingly}},
  \citenamefont {{McKenna}}, \citenamefont {{Pickles}}, \citenamefont
  {{Poznanski}}, \citenamefont {{Rahmer}}, \citenamefont {{Rau}}, \citenamefont
  {{Rosing}}, \citenamefont {{Shara}}, \citenamefont {{Smith}}, \citenamefont
  {{Starr}}, \citenamefont {{Sullivan}}, \citenamefont {{Velur}}, \citenamefont
  {{Walters}},\ and\ \citenamefont {{Zolkower}}}]{Law+2009_PTF}%
  \BibitemOpen
  \bibfield  {author} {\bibinfo {author} {\bibfnamefont {N.~M.}\ \bibnamefont
  {{Law}}}, \bibinfo {author} {\bibfnamefont {S.~R.}\ \bibnamefont
  {{Kulkarni}}}, \bibinfo {author} {\bibfnamefont {R.~G.}\ \bibnamefont
  {{Dekany}}}, \bibinfo {author} {\bibfnamefont {E.~O.}\ \bibnamefont
  {{Ofek}}}, \bibinfo {author} {\bibfnamefont {R.~M.}\ \bibnamefont
  {{Quimby}}}, \bibinfo {author} {\bibfnamefont {P.~E.}\ \bibnamefont
  {{Nugent}}}, \bibinfo {author} {\bibfnamefont {J.}~\bibnamefont {{Surace}}},
  \bibinfo {author} {\bibfnamefont {C.~C.}\ \bibnamefont {{Grillmair}}},
  \bibinfo {author} {\bibfnamefont {J.~S.}\ \bibnamefont {{Bloom}}}, \bibinfo
  {author} {\bibfnamefont {M.~M.}\ \bibnamefont {{Kasliwal}}}, \bibinfo
  {author} {\bibfnamefont {L.}~\bibnamefont {{Bildsten}}}, \bibinfo {author}
  {\bibfnamefont {T.}~\bibnamefont {{Brown}}}, \bibinfo {author} {\bibfnamefont
  {S.~B.}\ \bibnamefont {{Cenko}}}, \bibinfo {author} {\bibfnamefont
  {D.}~\bibnamefont {{Ciardi}}}, \bibinfo {author} {\bibfnamefont
  {E.}~\bibnamefont {{Croner}}}, \bibinfo {author} {\bibfnamefont {S.~G.}\
  \bibnamefont {{Djorgovski}}}, \bibinfo {author} {\bibfnamefont
  {J.}~\bibnamefont {{van Eyken}}}, \bibinfo {author} {\bibfnamefont {A.~V.}\
  \bibnamefont {{Filippenko}}}, \bibinfo {author} {\bibfnamefont {D.~B.}\
  \bibnamefont {{Fox}}}, \bibinfo {author} {\bibfnamefont {A.}~\bibnamefont
  {{Gal-Yam}}}, \bibinfo {author} {\bibfnamefont {D.}~\bibnamefont {{Hale}}},
  \bibinfo {author} {\bibfnamefont {N.}~\bibnamefont {{Hamam}}}, \bibinfo
  {author} {\bibfnamefont {G.}~\bibnamefont {{Helou}}}, \bibinfo {author}
  {\bibfnamefont {J.}~\bibnamefont {{Henning}}}, \bibinfo {author}
  {\bibfnamefont {D.~A.}\ \bibnamefont {{Howell}}}, \bibinfo {author}
  {\bibfnamefont {J.}~\bibnamefont {{Jacobsen}}}, \bibinfo {author}
  {\bibfnamefont {R.}~\bibnamefont {{Laher}}}, \bibinfo {author} {\bibfnamefont
  {S.}~\bibnamefont {{Mattingly}}}, \bibinfo {author} {\bibfnamefont
  {D.}~\bibnamefont {{McKenna}}}, \bibinfo {author} {\bibfnamefont
  {A.}~\bibnamefont {{Pickles}}}, \bibinfo {author} {\bibfnamefont
  {D.}~\bibnamefont {{Poznanski}}}, \bibinfo {author} {\bibfnamefont
  {G.}~\bibnamefont {{Rahmer}}}, \bibinfo {author} {\bibfnamefont
  {A.}~\bibnamefont {{Rau}}}, \bibinfo {author} {\bibfnamefont
  {W.}~\bibnamefont {{Rosing}}}, \bibinfo {author} {\bibfnamefont
  {M.}~\bibnamefont {{Shara}}}, \bibinfo {author} {\bibfnamefont
  {R.}~\bibnamefont {{Smith}}}, \bibinfo {author} {\bibfnamefont
  {D.}~\bibnamefont {{Starr}}}, \bibinfo {author} {\bibfnamefont
  {M.}~\bibnamefont {{Sullivan}}}, \bibinfo {author} {\bibfnamefont
  {V.}~\bibnamefont {{Velur}}}, \bibinfo {author} {\bibfnamefont
  {R.}~\bibnamefont {{Walters}}}, \ and\ \bibinfo {author} {\bibfnamefont
  {J.}~\bibnamefont {{Zolkower}}},\ }\href {\doibase 10.1086/648598} {\bibfield
   {journal} {\bibinfo  {journal} {\pasp}\ }\textbf {\bibinfo {volume} {121}},\
  \bibinfo {pages} {1395} (\bibinfo {year} {2009})},\ \Eprint
  {http://arxiv.org/abs/0906.5350} {arXiv:0906.5350 [astro-ph.IM]} \BibitemShut
  {NoStop}%
\bibitem [{\citenamefont {{Chambers}}\ \emph {et~al.}(2016)\citenamefont
  {{Chambers}}, \citenamefont {{Magnier}}, \citenamefont {{Metcalfe}},
  \citenamefont {{Flewelling}}, \citenamefont {{Huber}}, \citenamefont
  {{Waters}}, \citenamefont {{Denneau}}, \citenamefont {{Draper}},
  \citenamefont {{Farrow}}, \citenamefont {{Finkbeiner}}, \citenamefont
  {{Holmberg}}, \citenamefont {{Koppenhoefer}}, \citenamefont {{Price}},
  \citenamefont {{Rest}}, \citenamefont {{Saglia}}, \citenamefont {{Schlafly}},
  \citenamefont {{Smartt}}, \citenamefont {{Sweeney}}, \citenamefont
  {{Wainscoat}}, \citenamefont {{Burgett}}, \citenamefont {{Chastel}},
  \citenamefont {{Grav}}, \citenamefont {{Heasley}}, \citenamefont {{Hodapp}},
  \citenamefont {{Jedicke}}, \citenamefont {{Kaiser}}, \citenamefont
  {{Kudritzki}}, \citenamefont {{Luppino}}, \citenamefont {{Lupton}},
  \citenamefont {{Monet}}, \citenamefont {{Morgan}}, \citenamefont {{Onaka}},
  \citenamefont {{Shiao}}, \citenamefont {{Stubbs}}, \citenamefont {{Tonry}},
  \citenamefont {{White}}, \citenamefont {{Ba{\~n}ados}}, \citenamefont
  {{Bell}}, \citenamefont {{Bender}}, \citenamefont {{Bernard}}, \citenamefont
  {{Boegner}}, \citenamefont {{Boffi}}, \citenamefont {{Botticella}},
  \citenamefont {{Calamida}}, \citenamefont {{Casertano}}, \citenamefont
  {{Chen}}, \citenamefont {{Chen}}, \citenamefont {{Cole}}, \citenamefont
  {{Deacon}}, \citenamefont {{Frenk}}, \citenamefont {{Fitzsimmons}},
  \citenamefont {{Gezari}}, \citenamefont {{Gibbs}}, \citenamefont {{Goessl}},
  \citenamefont {{Goggia}}, \citenamefont {{Gourgue}}, \citenamefont
  {{Goldman}}, \citenamefont {{Grant}}, \citenamefont {{Grebel}}, \citenamefont
  {{Hambly}}, \citenamefont {{Hasinger}}, \citenamefont {{Heavens}},
  \citenamefont {{Heckman}}, \citenamefont {{Henderson}}, \citenamefont
  {{Henning}}, \citenamefont {{Holman}}, \citenamefont {{Hopp}}, \citenamefont
  {{Ip}}, \citenamefont {{Isani}}, \citenamefont {{Jackson}}, \citenamefont
  {{Keyes}}, \citenamefont {{Koekemoer}}, \citenamefont {{Kotak}},
  \citenamefont {{Le}}, \citenamefont {{Liska}}, \citenamefont {{Long}},
  \citenamefont {{Lucey}}, \citenamefont {{Liu}}, \citenamefont {{Martin}},
  \citenamefont {{Masci}}, \citenamefont {{McLean}}, \citenamefont {{Mindel}},
  \citenamefont {{Misra}}, \citenamefont {{Morganson}}, \citenamefont
  {{Murphy}}, \citenamefont {{Obaika}}, \citenamefont {{Narayan}},
  \citenamefont {{Nieto-Santisteban}}, \citenamefont {{Norberg}}, \citenamefont
  {{Peacock}}, \citenamefont {{Pier}}, \citenamefont {{Postman}}, \citenamefont
  {{Primak}}, \citenamefont {{Rae}}, \citenamefont {{Rai}}, \citenamefont
  {{Riess}}, \citenamefont {{Riffeser}}, \citenamefont {{Rix}}, \citenamefont
  {{R{\"o}ser}}, \citenamefont {{Russel}}, \citenamefont {{Rutz}},
  \citenamefont {{Schilbach}}, \citenamefont {{Schultz}}, \citenamefont
  {{Scolnic}}, \citenamefont {{Strolger}}, \citenamefont {{Szalay}},
  \citenamefont {{Seitz}}, \citenamefont {{Small}}, \citenamefont {{Smith}},
  \citenamefont {{Soderblom}}, \citenamefont {{Taylor}}, \citenamefont
  {{Thomson}}, \citenamefont {{Taylor}}, \citenamefont {{Thakar}},
  \citenamefont {{Thiel}}, \citenamefont {{Thilker}}, \citenamefont {{Unger}},
  \citenamefont {{Urata}}, \citenamefont {{Valenti}}, \citenamefont {{Wagner}},
  \citenamefont {{Walder}}, \citenamefont {{Walter}}, \citenamefont
  {{Watters}}, \citenamefont {{Werner}}, \citenamefont {{Wood-Vasey}},\ and\
  \citenamefont {{Wyse}}}]{Chambers+2016_PS1_Surveys}%
  \BibitemOpen
  \bibfield  {author} {\bibinfo {author} {\bibfnamefont {K.~C.}\ \bibnamefont
  {{Chambers}}}, \bibinfo {author} {\bibfnamefont {E.~A.}\ \bibnamefont
  {{Magnier}}}, \bibinfo {author} {\bibfnamefont {N.}~\bibnamefont
  {{Metcalfe}}}, \bibinfo {author} {\bibfnamefont {H.~A.}\ \bibnamefont
  {{Flewelling}}}, \bibinfo {author} {\bibfnamefont {M.~E.}\ \bibnamefont
  {{Huber}}}, \bibinfo {author} {\bibfnamefont {C.~Z.}\ \bibnamefont
  {{Waters}}}, \bibinfo {author} {\bibfnamefont {L.}~\bibnamefont {{Denneau}}},
  \bibinfo {author} {\bibfnamefont {P.~W.}\ \bibnamefont {{Draper}}}, \bibinfo
  {author} {\bibfnamefont {D.}~\bibnamefont {{Farrow}}}, \bibinfo {author}
  {\bibfnamefont {D.~P.}\ \bibnamefont {{Finkbeiner}}}, \bibinfo {author}
  {\bibfnamefont {C.}~\bibnamefont {{Holmberg}}}, \bibinfo {author}
  {\bibfnamefont {J.}~\bibnamefont {{Koppenhoefer}}}, \bibinfo {author}
  {\bibfnamefont {P.~A.}\ \bibnamefont {{Price}}}, \bibinfo {author}
  {\bibfnamefont {A.}~\bibnamefont {{Rest}}}, \bibinfo {author} {\bibfnamefont
  {R.~P.}\ \bibnamefont {{Saglia}}}, \bibinfo {author} {\bibfnamefont {E.~F.}\
  \bibnamefont {{Schlafly}}}, \bibinfo {author} {\bibfnamefont {S.~J.}\
  \bibnamefont {{Smartt}}}, \bibinfo {author} {\bibfnamefont {W.}~\bibnamefont
  {{Sweeney}}}, \bibinfo {author} {\bibfnamefont {R.~J.}\ \bibnamefont
  {{Wainscoat}}}, \bibinfo {author} {\bibfnamefont {W.~S.}\ \bibnamefont
  {{Burgett}}}, \bibinfo {author} {\bibfnamefont {S.}~\bibnamefont
  {{Chastel}}}, \bibinfo {author} {\bibfnamefont {T.}~\bibnamefont {{Grav}}},
  \bibinfo {author} {\bibfnamefont {J.~N.}\ \bibnamefont {{Heasley}}}, \bibinfo
  {author} {\bibfnamefont {K.~W.}\ \bibnamefont {{Hodapp}}}, \bibinfo {author}
  {\bibfnamefont {R.}~\bibnamefont {{Jedicke}}}, \bibinfo {author}
  {\bibfnamefont {N.}~\bibnamefont {{Kaiser}}}, \bibinfo {author}
  {\bibfnamefont {R.~P.}\ \bibnamefont {{Kudritzki}}}, \bibinfo {author}
  {\bibfnamefont {G.~A.}\ \bibnamefont {{Luppino}}}, \bibinfo {author}
  {\bibfnamefont {R.~H.}\ \bibnamefont {{Lupton}}}, \bibinfo {author}
  {\bibfnamefont {D.~G.}\ \bibnamefont {{Monet}}}, \bibinfo {author}
  {\bibfnamefont {J.~S.}\ \bibnamefont {{Morgan}}}, \bibinfo {author}
  {\bibfnamefont {P.~M.}\ \bibnamefont {{Onaka}}}, \bibinfo {author}
  {\bibfnamefont {B.}~\bibnamefont {{Shiao}}}, \bibinfo {author} {\bibfnamefont
  {C.~W.}\ \bibnamefont {{Stubbs}}}, \bibinfo {author} {\bibfnamefont {J.~L.}\
  \bibnamefont {{Tonry}}}, \bibinfo {author} {\bibfnamefont {R.}~\bibnamefont
  {{White}}}, \bibinfo {author} {\bibfnamefont {E.}~\bibnamefont
  {{Ba{\~n}ados}}}, \bibinfo {author} {\bibfnamefont {E.~F.}\ \bibnamefont
  {{Bell}}}, \bibinfo {author} {\bibfnamefont {R.}~\bibnamefont {{Bender}}},
  \bibinfo {author} {\bibfnamefont {E.~J.}\ \bibnamefont {{Bernard}}}, \bibinfo
  {author} {\bibfnamefont {M.}~\bibnamefont {{Boegner}}}, \bibinfo {author}
  {\bibfnamefont {F.}~\bibnamefont {{Boffi}}}, \bibinfo {author} {\bibfnamefont
  {M.~T.}\ \bibnamefont {{Botticella}}}, \bibinfo {author} {\bibfnamefont
  {A.}~\bibnamefont {{Calamida}}}, \bibinfo {author} {\bibfnamefont
  {S.}~\bibnamefont {{Casertano}}}, \bibinfo {author} {\bibfnamefont {W.~P.}\
  \bibnamefont {{Chen}}}, \bibinfo {author} {\bibfnamefont {X.}~\bibnamefont
  {{Chen}}}, \bibinfo {author} {\bibfnamefont {S.}~\bibnamefont {{Cole}}},
  \bibinfo {author} {\bibfnamefont {N.}~\bibnamefont {{Deacon}}}, \bibinfo
  {author} {\bibfnamefont {C.}~\bibnamefont {{Frenk}}}, \bibinfo {author}
  {\bibfnamefont {A.}~\bibnamefont {{Fitzsimmons}}}, \bibinfo {author}
  {\bibfnamefont {S.}~\bibnamefont {{Gezari}}}, \bibinfo {author}
  {\bibfnamefont {V.}~\bibnamefont {{Gibbs}}}, \bibinfo {author} {\bibfnamefont
  {C.}~\bibnamefont {{Goessl}}}, \bibinfo {author} {\bibfnamefont
  {T.}~\bibnamefont {{Goggia}}}, \bibinfo {author} {\bibfnamefont
  {R.}~\bibnamefont {{Gourgue}}}, \bibinfo {author} {\bibfnamefont
  {B.}~\bibnamefont {{Goldman}}}, \bibinfo {author} {\bibfnamefont
  {P.}~\bibnamefont {{Grant}}}, \bibinfo {author} {\bibfnamefont {E.~K.}\
  \bibnamefont {{Grebel}}}, \bibinfo {author} {\bibfnamefont {N.~C.}\
  \bibnamefont {{Hambly}}}, \bibinfo {author} {\bibfnamefont {G.}~\bibnamefont
  {{Hasinger}}}, \bibinfo {author} {\bibfnamefont {A.~F.}\ \bibnamefont
  {{Heavens}}}, \bibinfo {author} {\bibfnamefont {T.~M.}\ \bibnamefont
  {{Heckman}}}, \bibinfo {author} {\bibfnamefont {R.}~\bibnamefont
  {{Henderson}}}, \bibinfo {author} {\bibfnamefont {T.}~\bibnamefont
  {{Henning}}}, \bibinfo {author} {\bibfnamefont {M.}~\bibnamefont {{Holman}}},
  \bibinfo {author} {\bibfnamefont {U.}~\bibnamefont {{Hopp}}}, \bibinfo
  {author} {\bibfnamefont {W.~H.}\ \bibnamefont {{Ip}}}, \bibinfo {author}
  {\bibfnamefont {S.}~\bibnamefont {{Isani}}}, \bibinfo {author} {\bibfnamefont
  {M.}~\bibnamefont {{Jackson}}}, \bibinfo {author} {\bibfnamefont {C.~D.}\
  \bibnamefont {{Keyes}}}, \bibinfo {author} {\bibfnamefont {A.~M.}\
  \bibnamefont {{Koekemoer}}}, \bibinfo {author} {\bibfnamefont
  {R.}~\bibnamefont {{Kotak}}}, \bibinfo {author} {\bibfnamefont
  {D.}~\bibnamefont {{Le}}}, \bibinfo {author} {\bibfnamefont {D.}~\bibnamefont
  {{Liska}}}, \bibinfo {author} {\bibfnamefont {K.~S.}\ \bibnamefont {{Long}}},
  \bibinfo {author} {\bibfnamefont {J.~R.}\ \bibnamefont {{Lucey}}}, \bibinfo
  {author} {\bibfnamefont {M.}~\bibnamefont {{Liu}}}, \bibinfo {author}
  {\bibfnamefont {N.~F.}\ \bibnamefont {{Martin}}}, \bibinfo {author}
  {\bibfnamefont {G.}~\bibnamefont {{Masci}}}, \bibinfo {author} {\bibfnamefont
  {B.}~\bibnamefont {{McLean}}}, \bibinfo {author} {\bibfnamefont
  {E.}~\bibnamefont {{Mindel}}}, \bibinfo {author} {\bibfnamefont
  {P.}~\bibnamefont {{Misra}}}, \bibinfo {author} {\bibfnamefont
  {E.}~\bibnamefont {{Morganson}}}, \bibinfo {author} {\bibfnamefont
  {D.~N.~A.}\ \bibnamefont {{Murphy}}}, \bibinfo {author} {\bibfnamefont
  {A.}~\bibnamefont {{Obaika}}}, \bibinfo {author} {\bibfnamefont
  {G.}~\bibnamefont {{Narayan}}}, \bibinfo {author} {\bibfnamefont {M.~A.}\
  \bibnamefont {{Nieto-Santisteban}}}, \bibinfo {author} {\bibfnamefont
  {P.}~\bibnamefont {{Norberg}}}, \bibinfo {author} {\bibfnamefont {J.~A.}\
  \bibnamefont {{Peacock}}}, \bibinfo {author} {\bibfnamefont {E.~A.}\
  \bibnamefont {{Pier}}}, \bibinfo {author} {\bibfnamefont {M.}~\bibnamefont
  {{Postman}}}, \bibinfo {author} {\bibfnamefont {N.}~\bibnamefont {{Primak}}},
  \bibinfo {author} {\bibfnamefont {C.}~\bibnamefont {{Rae}}}, \bibinfo
  {author} {\bibfnamefont {A.}~\bibnamefont {{Rai}}}, \bibinfo {author}
  {\bibfnamefont {A.}~\bibnamefont {{Riess}}}, \bibinfo {author} {\bibfnamefont
  {A.}~\bibnamefont {{Riffeser}}}, \bibinfo {author} {\bibfnamefont {H.~W.}\
  \bibnamefont {{Rix}}}, \bibinfo {author} {\bibfnamefont {S.}~\bibnamefont
  {{R{\"o}ser}}}, \bibinfo {author} {\bibfnamefont {R.}~\bibnamefont
  {{Russel}}}, \bibinfo {author} {\bibfnamefont {L.}~\bibnamefont {{Rutz}}},
  \bibinfo {author} {\bibfnamefont {E.}~\bibnamefont {{Schilbach}}}, \bibinfo
  {author} {\bibfnamefont {A.~S.~B.}\ \bibnamefont {{Schultz}}}, \bibinfo
  {author} {\bibfnamefont {D.}~\bibnamefont {{Scolnic}}}, \bibinfo {author}
  {\bibfnamefont {L.}~\bibnamefont {{Strolger}}}, \bibinfo {author}
  {\bibfnamefont {A.}~\bibnamefont {{Szalay}}}, \bibinfo {author}
  {\bibfnamefont {S.}~\bibnamefont {{Seitz}}}, \bibinfo {author} {\bibfnamefont
  {E.}~\bibnamefont {{Small}}}, \bibinfo {author} {\bibfnamefont {K.~W.}\
  \bibnamefont {{Smith}}}, \bibinfo {author} {\bibfnamefont {D.~R.}\
  \bibnamefont {{Soderblom}}}, \bibinfo {author} {\bibfnamefont
  {P.}~\bibnamefont {{Taylor}}}, \bibinfo {author} {\bibfnamefont
  {R.}~\bibnamefont {{Thomson}}}, \bibinfo {author} {\bibfnamefont {A.~N.}\
  \bibnamefont {{Taylor}}}, \bibinfo {author} {\bibfnamefont {A.~R.}\
  \bibnamefont {{Thakar}}}, \bibinfo {author} {\bibfnamefont {J.}~\bibnamefont
  {{Thiel}}}, \bibinfo {author} {\bibfnamefont {D.}~\bibnamefont {{Thilker}}},
  \bibinfo {author} {\bibfnamefont {D.}~\bibnamefont {{Unger}}}, \bibinfo
  {author} {\bibfnamefont {Y.}~\bibnamefont {{Urata}}}, \bibinfo {author}
  {\bibfnamefont {J.}~\bibnamefont {{Valenti}}}, \bibinfo {author}
  {\bibfnamefont {J.}~\bibnamefont {{Wagner}}}, \bibinfo {author}
  {\bibfnamefont {T.}~\bibnamefont {{Walder}}}, \bibinfo {author}
  {\bibfnamefont {F.}~\bibnamefont {{Walter}}}, \bibinfo {author}
  {\bibfnamefont {S.~P.}\ \bibnamefont {{Watters}}}, \bibinfo {author}
  {\bibfnamefont {S.}~\bibnamefont {{Werner}}}, \bibinfo {author}
  {\bibfnamefont {W.~M.}\ \bibnamefont {{Wood-Vasey}}}, \ and\ \bibinfo
  {author} {\bibfnamefont {R.}~\bibnamefont {{Wyse}}},\ }\href@noop {}
  {\bibfield  {journal} {\bibinfo  {journal} {arXiv e-prints}\ ,\ \bibinfo
  {eid} {arXiv:1612.05560}} (\bibinfo {year} {2016})},\ \Eprint
  {http://arxiv.org/abs/1612.05560} {arXiv:1612.05560 [astro-ph.IM]}
  \BibitemShut {NoStop}%
\bibitem [{\citenamefont {{Gorbovskoy}}\ \emph {et~al.}(2013)\citenamefont
  {{Gorbovskoy}}, \citenamefont {{Lipunov}}, \citenamefont {{Kornilov}},
  \citenamefont {{Belinski}}, \citenamefont {{Kuvshinov}}, \citenamefont
  {{Tyurina}}, \citenamefont {{Sankovich}}, \citenamefont {{Krylov}},
  \citenamefont {{Shatskiy}}, \citenamefont {{Balanutsa}}, \citenamefont
  {{Chazov}}, \citenamefont {{Kuznetsov}}, \citenamefont {{Zimnukhov}},
  \citenamefont {{Shumkov}}, \citenamefont {{Shurpakov}}, \citenamefont
  {{Senik}}, \citenamefont {{Gareeva}}, \citenamefont {{Pruzhinskaya}},
  \citenamefont {{Tlatov}}, \citenamefont {{Parkhomenko}}, \citenamefont
  {{Dormidontov}}, \citenamefont {{Krushinsky}}, \citenamefont {{Punanova}},
  \citenamefont {{Zalozhnyh}}, \citenamefont {{Popov}}, \citenamefont
  {{Burdanov}}, \citenamefont {{Yazev}}, \citenamefont {{Budnev}},
  \citenamefont {{Ivanov}}, \citenamefont {{Konstantinov}}, \citenamefont
  {{Gress}}, \citenamefont {{Chuvalaev}}, \citenamefont {{Yurkov}},
  \citenamefont {{Sergienko}}, \citenamefont {{Kudelina}}, \citenamefont
  {{Sinyakov}}, \citenamefont {{Karachentsev}}, \citenamefont {{Moiseev}},\
  and\ \citenamefont
  {{Fatkhullin}}}]{Gorbovskoy+2013_MASTER_OpticalTelescopes}%
  \BibitemOpen
  \bibfield  {author} {\bibinfo {author} {\bibfnamefont {E.~S.}\ \bibnamefont
  {{Gorbovskoy}}}, \bibinfo {author} {\bibfnamefont {V.~M.}\ \bibnamefont
  {{Lipunov}}}, \bibinfo {author} {\bibfnamefont {V.~G.}\ \bibnamefont
  {{Kornilov}}}, \bibinfo {author} {\bibfnamefont {A.~A.}\ \bibnamefont
  {{Belinski}}}, \bibinfo {author} {\bibfnamefont {D.~A.}\ \bibnamefont
  {{Kuvshinov}}}, \bibinfo {author} {\bibfnamefont {N.~V.}\ \bibnamefont
  {{Tyurina}}}, \bibinfo {author} {\bibfnamefont {A.~V.}\ \bibnamefont
  {{Sankovich}}}, \bibinfo {author} {\bibfnamefont {A.~V.}\ \bibnamefont
  {{Krylov}}}, \bibinfo {author} {\bibfnamefont {N.~I.}\ \bibnamefont
  {{Shatskiy}}}, \bibinfo {author} {\bibfnamefont {P.~V.}\ \bibnamefont
  {{Balanutsa}}}, \bibinfo {author} {\bibfnamefont {V.~V.}\ \bibnamefont
  {{Chazov}}}, \bibinfo {author} {\bibfnamefont {A.~S.}\ \bibnamefont
  {{Kuznetsov}}}, \bibinfo {author} {\bibfnamefont {A.~S.}\ \bibnamefont
  {{Zimnukhov}}}, \bibinfo {author} {\bibfnamefont {V.~P.}\ \bibnamefont
  {{Shumkov}}}, \bibinfo {author} {\bibfnamefont {S.~E.}\ \bibnamefont
  {{Shurpakov}}}, \bibinfo {author} {\bibfnamefont {V.~A.}\ \bibnamefont
  {{Senik}}}, \bibinfo {author} {\bibfnamefont {D.~V.}\ \bibnamefont
  {{Gareeva}}}, \bibinfo {author} {\bibfnamefont {M.~V.}\ \bibnamefont
  {{Pruzhinskaya}}}, \bibinfo {author} {\bibfnamefont {A.~G.}\ \bibnamefont
  {{Tlatov}}}, \bibinfo {author} {\bibfnamefont {A.~V.}\ \bibnamefont
  {{Parkhomenko}}}, \bibinfo {author} {\bibfnamefont {D.~V.}\ \bibnamefont
  {{Dormidontov}}}, \bibinfo {author} {\bibfnamefont {V.~V.}\ \bibnamefont
  {{Krushinsky}}}, \bibinfo {author} {\bibfnamefont {A.~F.}\ \bibnamefont
  {{Punanova}}}, \bibinfo {author} {\bibfnamefont {I.~S.}\ \bibnamefont
  {{Zalozhnyh}}}, \bibinfo {author} {\bibfnamefont {A.~A.}\ \bibnamefont
  {{Popov}}}, \bibinfo {author} {\bibfnamefont {A.~Y.}\ \bibnamefont
  {{Burdanov}}}, \bibinfo {author} {\bibfnamefont {S.~A.}\ \bibnamefont
  {{Yazev}}}, \bibinfo {author} {\bibfnamefont {N.~M.}\ \bibnamefont
  {{Budnev}}}, \bibinfo {author} {\bibfnamefont {K.~I.}\ \bibnamefont
  {{Ivanov}}}, \bibinfo {author} {\bibfnamefont {E.~N.}\ \bibnamefont
  {{Konstantinov}}}, \bibinfo {author} {\bibfnamefont {O.~A.}\ \bibnamefont
  {{Gress}}}, \bibinfo {author} {\bibfnamefont {O.~V.}\ \bibnamefont
  {{Chuvalaev}}}, \bibinfo {author} {\bibfnamefont {V.~V.}\ \bibnamefont
  {{Yurkov}}}, \bibinfo {author} {\bibfnamefont {Y.~P.}\ \bibnamefont
  {{Sergienko}}}, \bibinfo {author} {\bibfnamefont {I.~V.}\ \bibnamefont
  {{Kudelina}}}, \bibinfo {author} {\bibfnamefont {E.~V.}\ \bibnamefont
  {{Sinyakov}}}, \bibinfo {author} {\bibfnamefont {I.~D.}\ \bibnamefont
  {{Karachentsev}}}, \bibinfo {author} {\bibfnamefont {A.~V.}\ \bibnamefont
  {{Moiseev}}}, \ and\ \bibinfo {author} {\bibfnamefont {T.~A.}\ \bibnamefont
  {{Fatkhullin}}},\ }\href {\doibase 10.1134/S1063772913040033} {\bibfield
  {journal} {\bibinfo  {journal} {Astronomy Reports}\ }\textbf {\bibinfo
  {volume} {57}},\ \bibinfo {pages} {233} (\bibinfo {year} {2013})},\ \Eprint
  {http://arxiv.org/abs/1305.1620} {arXiv:1305.1620 [astro-ph.HE]} \BibitemShut
  {NoStop}%
\bibitem [{\citenamefont {{Kochanek}}\ \emph {et~al.}(2017)\citenamefont
  {{Kochanek}}, \citenamefont {{Shappee}}, \citenamefont {{Stanek}},
  \citenamefont {{Holoien}}, \citenamefont {{Thompson}}, \citenamefont
  {{Prieto}}, \citenamefont {{Dong}}, \citenamefont {{Shields}}, \citenamefont
  {{Will}}, \citenamefont {{Britt}}, \citenamefont {{Perzanowski}},\ and\
  \citenamefont {{Pojma{\'n}ski}}}]{Kochanek+2017_ASASSN_VarStars}%
  \BibitemOpen
  \bibfield  {author} {\bibinfo {author} {\bibfnamefont {C.~S.}\ \bibnamefont
  {{Kochanek}}}, \bibinfo {author} {\bibfnamefont {B.~J.}\ \bibnamefont
  {{Shappee}}}, \bibinfo {author} {\bibfnamefont {K.~Z.}\ \bibnamefont
  {{Stanek}}}, \bibinfo {author} {\bibfnamefont {T.~W.~S.}\ \bibnamefont
  {{Holoien}}}, \bibinfo {author} {\bibfnamefont {T.~A.}\ \bibnamefont
  {{Thompson}}}, \bibinfo {author} {\bibfnamefont {J.~L.}\ \bibnamefont
  {{Prieto}}}, \bibinfo {author} {\bibfnamefont {S.}~\bibnamefont {{Dong}}},
  \bibinfo {author} {\bibfnamefont {J.~V.}\ \bibnamefont {{Shields}}}, \bibinfo
  {author} {\bibfnamefont {D.}~\bibnamefont {{Will}}}, \bibinfo {author}
  {\bibfnamefont {C.}~\bibnamefont {{Britt}}}, \bibinfo {author} {\bibfnamefont
  {D.}~\bibnamefont {{Perzanowski}}}, \ and\ \bibinfo {author} {\bibfnamefont
  {G.}~\bibnamefont {{Pojma{\'n}ski}}},\ }\href {\doibase
  10.1088/1538-3873/aa80d9} {\bibfield  {journal} {\bibinfo  {journal} {\pasp}\
  }\textbf {\bibinfo {volume} {129}},\ \bibinfo {pages} {104502} (\bibinfo
  {year} {2017})},\ \Eprint {http://arxiv.org/abs/1706.07060} {arXiv:1706.07060
  [astro-ph.SR]} \BibitemShut {NoStop}%
\bibitem [{\citenamefont {{Dyer}}\ \emph {et~al.}(2018)\citenamefont {{Dyer}},
  \citenamefont {{Dhillon}}, \citenamefont {{Littlefair}}, \citenamefont
  {{Steeghs}}, \citenamefont {{Ulaczyk}}, \citenamefont {{Chote}},
  \citenamefont {{Galloway}},\ and\ \citenamefont
  {{Rol}}}]{Martin+2018_GOTO_Control}%
  \BibitemOpen
  \bibfield  {author} {\bibinfo {author} {\bibfnamefont {M.~J.}\ \bibnamefont
  {{Dyer}}}, \bibinfo {author} {\bibfnamefont {V.~S.}\ \bibnamefont
  {{Dhillon}}}, \bibinfo {author} {\bibfnamefont {S.}~\bibnamefont
  {{Littlefair}}}, \bibinfo {author} {\bibfnamefont {D.}~\bibnamefont
  {{Steeghs}}}, \bibinfo {author} {\bibfnamefont {K.}~\bibnamefont
  {{Ulaczyk}}}, \bibinfo {author} {\bibfnamefont {P.}~\bibnamefont {{Chote}}},
  \bibinfo {author} {\bibfnamefont {D.}~\bibnamefont {{Galloway}}}, \ and\
  \bibinfo {author} {\bibfnamefont {E.}~\bibnamefont {{Rol}}},\ }in\ \href
  {\doibase 10.1117/12.2311865} {\emph {\bibinfo {booktitle} {\procspie}}},\
  \bibinfo {series} {Society of Photo-Optical Instrumentation Engineers (SPIE)
  Conference Series}, Vol.\ \bibinfo {volume} {10704}\ (\bibinfo {year}
  {2018})\ p.\ \bibinfo {pages} {107040C},\ \Eprint
  {http://arxiv.org/abs/1807.01614} {arXiv:1807.01614 [astro-ph.IM]}
  \BibitemShut {NoStop}%
\bibitem [{\citenamefont {{Bellm}}\ \emph
  {et~al.}(2019{\natexlab{a}})\citenamefont {{Bellm}}, \citenamefont
  {{Kulkarni}}, \citenamefont {{Graham}}, \citenamefont {{Dekany}},
  \citenamefont {{Smith}}, \citenamefont {{Riddle}}, \citenamefont {{Masci}},
  \citenamefont {{Helou}}, \citenamefont {{Prince}},\ and\ \citenamefont
  {{Adams}}}]{Bellm+2019_ZTF_Overview}%
  \BibitemOpen
  \bibfield  {author} {\bibinfo {author} {\bibfnamefont {E.~C.}\ \bibnamefont
  {{Bellm}}}, \bibinfo {author} {\bibfnamefont {S.~R.}\ \bibnamefont
  {{Kulkarni}}}, \bibinfo {author} {\bibfnamefont {M.~J.}\ \bibnamefont
  {{Graham}}}, \bibinfo {author} {\bibfnamefont {R.}~\bibnamefont {{Dekany}}},
  \bibinfo {author} {\bibfnamefont {R.~M.}\ \bibnamefont {{Smith}}}, \bibinfo
  {author} {\bibfnamefont {R.}~\bibnamefont {{Riddle}}}, \bibinfo {author}
  {\bibfnamefont {F.~J.}\ \bibnamefont {{Masci}}}, \bibinfo {author}
  {\bibfnamefont {G.}~\bibnamefont {{Helou}}}, \bibinfo {author} {\bibfnamefont
  {T.~A.}\ \bibnamefont {{Prince}}}, \ and\ \bibinfo {author} {\bibfnamefont
  {S.~M.}\ \bibnamefont {{Adams}}},\ }\href {\doibase 10.1088/1538-3873/aaecbe}
  {\bibfield  {journal} {\bibinfo  {journal} {\pasp}\ }\textbf {\bibinfo
  {volume} {131}},\ \bibinfo {pages} {018002} (\bibinfo {year}
  {2019}{\natexlab{a}})},\ \Eprint {http://arxiv.org/abs/1902.01932}
  {arXiv:1902.01932 [astro-ph.IM]} \BibitemShut {NoStop}%
\bibitem [{\citenamefont {{Tyson}}\ \emph {et~al.}(2001)\citenamefont
  {{Tyson}}, \citenamefont {{Wittman}},\ and\ \citenamefont
  {{Angel}}}]{Tyson+2001_LSST}%
  \BibitemOpen
  \bibfield  {author} {\bibinfo {author} {\bibfnamefont {J.~A.}\ \bibnamefont
  {{Tyson}}}, \bibinfo {author} {\bibfnamefont {D.~M.}\ \bibnamefont
  {{Wittman}}}, \ and\ \bibinfo {author} {\bibfnamefont {J.~R.~P.}\
  \bibnamefont {{Angel}}},\ }in\ \href@noop {} {\emph {\bibinfo {booktitle}
  {Gravitational Lensing: Recent Progress and Future Go}}},\ \bibinfo {series}
  {Astronomical Society of the Pacific Conference Series}, Vol.\ \bibinfo
  {volume} {237},\ \bibinfo {editor} {edited by\ \bibinfo {editor}
  {\bibfnamefont {T.~G.}\ \bibnamefont {{Brainerd}}}\ and\ \bibinfo {editor}
  {\bibfnamefont {C.~S.}\ \bibnamefont {{Kochanek}}}}\ (\bibinfo {year}
  {2001})\ p.\ \bibinfo {pages} {417},\ \Eprint
  {http://arxiv.org/abs/astro-ph/0005381} {arXiv:astro-ph/0005381 [astro-ph]}
  \BibitemShut {NoStop}%
\bibitem [{\citenamefont {{Abbott}}\ \emph {et~al.}(2017)\citenamefont
  {{Abbott}}, \citenamefont {{Abbott}}, \citenamefont {{Abbott}}, \citenamefont
  {{Acernese}}, \citenamefont {{Ackley}}, \citenamefont {{Adams}},
  \citenamefont {{Adams}}, \citenamefont {{Addesso}}, \citenamefont
  {{Adhikari}},\ and\ \citenamefont {{Adya}}}]{Abbott+2017_GW170817_LIGO}%
  \BibitemOpen
  \bibfield  {author} {\bibinfo {author} {\bibfnamefont {B.~P.}\ \bibnamefont
  {{Abbott}}}, \bibinfo {author} {\bibfnamefont {R.}~\bibnamefont {{Abbott}}},
  \bibinfo {author} {\bibfnamefont {T.~D.}\ \bibnamefont {{Abbott}}}, \bibinfo
  {author} {\bibfnamefont {F.}~\bibnamefont {{Acernese}}}, \bibinfo {author}
  {\bibfnamefont {K.}~\bibnamefont {{Ackley}}}, \bibinfo {author}
  {\bibfnamefont {C.}~\bibnamefont {{Adams}}}, \bibinfo {author} {\bibfnamefont
  {T.}~\bibnamefont {{Adams}}}, \bibinfo {author} {\bibfnamefont
  {P.}~\bibnamefont {{Addesso}}}, \bibinfo {author} {\bibfnamefont {R.~X.}\
  \bibnamefont {{Adhikari}}}, \ and\ \bibinfo {author} {\bibfnamefont {V.~B.}\
  \bibnamefont {{Adya}}},\ }\href {\doibase 10.1103/PhysRevLett.119.161101}
  {\bibfield  {journal} {\bibinfo  {journal} {\prl}\ }\textbf {\bibinfo
  {volume} {119}},\ \bibinfo {eid} {161101} (\bibinfo {year} {2017})},\ \Eprint
  {http://arxiv.org/abs/1710.05832} {arXiv:1710.05832 [gr-qc]} \BibitemShut
  {NoStop}%
\bibitem [{\citenamefont {{Cenko}}\ \emph {et~al.}(2013)\citenamefont
  {{Cenko}}, \citenamefont {{Kulkarni}}, \citenamefont {{Horesh}},
  \citenamefont {{Corsi}}, \citenamefont {{Fox}}, \citenamefont {{Carpenter}},
  \citenamefont {{Frail}}, \citenamefont {{Nugent}}, \citenamefont {{Perley}},\
  and\ \citenamefont {{Gruber}}}]{Cenko+2013_PTF11agg}%
  \BibitemOpen
  \bibfield  {author} {\bibinfo {author} {\bibfnamefont {S.~B.}\ \bibnamefont
  {{Cenko}}}, \bibinfo {author} {\bibfnamefont {S.~R.}\ \bibnamefont
  {{Kulkarni}}}, \bibinfo {author} {\bibfnamefont {A.}~\bibnamefont
  {{Horesh}}}, \bibinfo {author} {\bibfnamefont {A.}~\bibnamefont {{Corsi}}},
  \bibinfo {author} {\bibfnamefont {D.~B.}\ \bibnamefont {{Fox}}}, \bibinfo
  {author} {\bibfnamefont {J.}~\bibnamefont {{Carpenter}}}, \bibinfo {author}
  {\bibfnamefont {D.~A.}\ \bibnamefont {{Frail}}}, \bibinfo {author}
  {\bibfnamefont {P.~E.}\ \bibnamefont {{Nugent}}}, \bibinfo {author}
  {\bibfnamefont {D.~A.}\ \bibnamefont {{Perley}}}, \ and\ \bibinfo {author}
  {\bibfnamefont {D.}~\bibnamefont {{Gruber}}},\ }\href {\doibase
  10.1088/0004-637X/769/2/130} {\bibfield  {journal} {\bibinfo  {journal}
  {\apj}\ }\textbf {\bibinfo {volume} {769}},\ \bibinfo {eid} {130} (\bibinfo
  {year} {2013})},\ \Eprint {http://arxiv.org/abs/1304.4236} {arXiv:1304.4236
  [astro-ph.CO]} \BibitemShut {NoStop}%
\bibitem [{\citenamefont {{Charbonneau}}\ \emph {et~al.}(2000)\citenamefont
  {{Charbonneau}}, \citenamefont {{Brown}}, \citenamefont {{Latham}},\ and\
  \citenamefont {{Mayor}}}]{Charbonneau+2000_FirstTransitingExoplanet}%
  \BibitemOpen
  \bibfield  {author} {\bibinfo {author} {\bibfnamefont {D.}~\bibnamefont
  {{Charbonneau}}}, \bibinfo {author} {\bibfnamefont {T.~M.}\ \bibnamefont
  {{Brown}}}, \bibinfo {author} {\bibfnamefont {D.~W.}\ \bibnamefont
  {{Latham}}}, \ and\ \bibinfo {author} {\bibfnamefont {M.}~\bibnamefont
  {{Mayor}}},\ }\href {\doibase 10.1086/312457} {\bibfield  {journal} {\bibinfo
   {journal} {\apjl}\ }\textbf {\bibinfo {volume} {529}},\ \bibinfo {pages}
  {L45} (\bibinfo {year} {2000})},\ \Eprint
  {http://arxiv.org/abs/astro-ph/9911436} {arXiv:astro-ph/9911436 [astro-ph]}
  \BibitemShut {NoStop}%
\bibitem [{\citenamefont {{Nutzman}}\ and\ \citenamefont
  {{Charbonneau}}(2008)}]{Nutzman+Charbonneau2008_MEarth_DesignConsiderations}%
  \BibitemOpen
  \bibfield  {author} {\bibinfo {author} {\bibfnamefont {P.}~\bibnamefont
  {{Nutzman}}}\ and\ \bibinfo {author} {\bibfnamefont {D.}~\bibnamefont
  {{Charbonneau}}},\ }\href {\doibase 10.1086/533420} {\bibfield  {journal}
  {\bibinfo  {journal} {\pasp}\ }\textbf {\bibinfo {volume} {120}},\ \bibinfo
  {pages} {317} (\bibinfo {year} {2008})},\ \Eprint
  {http://arxiv.org/abs/0709.2879} {arXiv:0709.2879 [astro-ph]} \BibitemShut
  {NoStop}%
\bibitem [{\citenamefont {{Swift}}\ \emph {et~al.}(2015)\citenamefont
  {{Swift}}, \citenamefont {{Bottom}}, \citenamefont {{Johnson}}, \citenamefont
  {{Wright}}, \citenamefont {{McCrady}}, \citenamefont {{Wittenmyer}},
  \citenamefont {{Plavchan}}, \citenamefont {{Riddle}}, \citenamefont
  {{Muirhead}}, \citenamefont {{Herzig}}, \citenamefont {{Myles}},
  \citenamefont {{Blake}}, \citenamefont {{Eastman}}, \citenamefont {{Beatty}},
  \citenamefont {{Barnes}}, \citenamefont {{Gibson}}, \citenamefont {{Lin}},
  \citenamefont {{Zhao}}, \citenamefont {{Gardner}}, \citenamefont {{Falco}},
  \citenamefont {{Criswell}}, \citenamefont {{Nava}}, \citenamefont
  {{Robinson}}, \citenamefont {{Sliski}}, \citenamefont {{Hedrick}},
  \citenamefont {{Ivarsen}}, \citenamefont {{Hjelstrom}}, \citenamefont {{de
  Vera}},\ and\ \citenamefont
  {{Szentgyorgyi}}}]{Swift+2015_Minerva_ExoPlanets_DesignCommissioning}%
  \BibitemOpen
  \bibfield  {author} {\bibinfo {author} {\bibfnamefont {J.~J.}\ \bibnamefont
  {{Swift}}}, \bibinfo {author} {\bibfnamefont {M.}~\bibnamefont {{Bottom}}},
  \bibinfo {author} {\bibfnamefont {J.~A.}\ \bibnamefont {{Johnson}}}, \bibinfo
  {author} {\bibfnamefont {J.~T.}\ \bibnamefont {{Wright}}}, \bibinfo {author}
  {\bibfnamefont {N.}~\bibnamefont {{McCrady}}}, \bibinfo {author}
  {\bibfnamefont {R.~A.}\ \bibnamefont {{Wittenmyer}}}, \bibinfo {author}
  {\bibfnamefont {P.}~\bibnamefont {{Plavchan}}}, \bibinfo {author}
  {\bibfnamefont {R.}~\bibnamefont {{Riddle}}}, \bibinfo {author}
  {\bibfnamefont {P.~S.}\ \bibnamefont {{Muirhead}}}, \bibinfo {author}
  {\bibfnamefont {E.}~\bibnamefont {{Herzig}}}, \bibinfo {author}
  {\bibfnamefont {J.}~\bibnamefont {{Myles}}}, \bibinfo {author} {\bibfnamefont
  {C.~H.}\ \bibnamefont {{Blake}}}, \bibinfo {author} {\bibfnamefont
  {J.}~\bibnamefont {{Eastman}}}, \bibinfo {author} {\bibfnamefont {T.~G.}\
  \bibnamefont {{Beatty}}}, \bibinfo {author} {\bibfnamefont {S.~I.}\
  \bibnamefont {{Barnes}}}, \bibinfo {author} {\bibfnamefont {S.~R.}\
  \bibnamefont {{Gibson}}}, \bibinfo {author} {\bibfnamefont {B.}~\bibnamefont
  {{Lin}}}, \bibinfo {author} {\bibfnamefont {M.}~\bibnamefont {{Zhao}}},
  \bibinfo {author} {\bibfnamefont {P.}~\bibnamefont {{Gardner}}}, \bibinfo
  {author} {\bibfnamefont {E.}~\bibnamefont {{Falco}}}, \bibinfo {author}
  {\bibfnamefont {S.}~\bibnamefont {{Criswell}}}, \bibinfo {author}
  {\bibfnamefont {C.}~\bibnamefont {{Nava}}}, \bibinfo {author} {\bibfnamefont
  {C.}~\bibnamefont {{Robinson}}}, \bibinfo {author} {\bibfnamefont {D.~H.}\
  \bibnamefont {{Sliski}}}, \bibinfo {author} {\bibfnamefont {R.}~\bibnamefont
  {{Hedrick}}}, \bibinfo {author} {\bibfnamefont {K.}~\bibnamefont
  {{Ivarsen}}}, \bibinfo {author} {\bibfnamefont {A.}~\bibnamefont
  {{Hjelstrom}}}, \bibinfo {author} {\bibfnamefont {J.}~\bibnamefont {{de
  Vera}}}, \ and\ \bibinfo {author} {\bibfnamefont {A.}~\bibnamefont
  {{Szentgyorgyi}}},\ }\href {\doibase 10.1117/1.JATIS.1.2.027002} {\bibfield
  {journal} {\bibinfo  {journal} {Journal of Astronomical Telescopes,
  Instruments, and Systems}\ }\textbf {\bibinfo {volume} {1}},\ \bibinfo {eid}
  {027002} (\bibinfo {year} {2015})},\ \Eprint {http://arxiv.org/abs/1411.3724}
  {arXiv:1411.3724 [astro-ph.IM]} \BibitemShut {NoStop}%
\bibitem [{\citenamefont {{Tonry}}(2011)}]{Tonry2011_ATLAS_SurveyCapability}%
  \BibitemOpen
  \bibfield  {author} {\bibinfo {author} {\bibfnamefont {J.~L.}\ \bibnamefont
  {{Tonry}}},\ }\href {\doibase 10.1086/657997} {\bibfield  {journal} {\bibinfo
   {journal} {\pasp}\ }\textbf {\bibinfo {volume} {123}},\ \bibinfo {pages}
  {58} (\bibinfo {year} {2011})},\ \Eprint {http://arxiv.org/abs/1011.1028}
  {arXiv:1011.1028 [astro-ph.IM]} \BibitemShut {NoStop}%
\bibitem [{\citenamefont {{Djorgovski}}\ \emph {et~al.}(2012)\citenamefont
  {{Djorgovski}}, \citenamefont {{Mahabal}}, \citenamefont {{Drake}},
  \citenamefont {{Graham}}, \citenamefont {{Donalek}},\ and\ \citenamefont
  {{Williams}}}]{Djorgovski2012_SkySurveysReview}%
  \BibitemOpen
  \bibfield  {author} {\bibinfo {author} {\bibfnamefont {S.~G.}\ \bibnamefont
  {{Djorgovski}}}, \bibinfo {author} {\bibfnamefont {A.~A.}\ \bibnamefont
  {{Mahabal}}}, \bibinfo {author} {\bibfnamefont {A.~J.}\ \bibnamefont
  {{Drake}}}, \bibinfo {author} {\bibfnamefont {M.~J.}\ \bibnamefont
  {{Graham}}}, \bibinfo {author} {\bibfnamefont {C.}~\bibnamefont {{Donalek}}},
  \ and\ \bibinfo {author} {\bibfnamefont {R.}~\bibnamefont {{Williams}}},\
  }in\ \href {\doibase 10.1017/S1743921312000488} {\emph {\bibinfo {booktitle}
  {New Horizons in Time Domain Astronomy}}},\ \bibinfo {series} {IAU
  Symposium}, Vol.\ \bibinfo {volume} {285},\ \bibinfo {editor} {edited by\
  \bibinfo {editor} {\bibfnamefont {E.}~\bibnamefont {{Griffin}}}, \bibinfo
  {editor} {\bibfnamefont {R.}~\bibnamefont {{Hanisch}}}, \ and\ \bibinfo
  {editor} {\bibfnamefont {R.}~\bibnamefont {{Seaman}}}}\ (\bibinfo {year}
  {2012})\ pp.\ \bibinfo {pages} {141--146},\ \Eprint
  {http://arxiv.org/abs/1111.2078} {arXiv:1111.2078 [astro-ph.IM]} \BibitemShut
  {NoStop}%
\bibitem [{\citenamefont {{Pepper}}\ \emph {et~al.}(2002)\citenamefont
  {{Pepper}}, \citenamefont {{Gould}},\ and\ \citenamefont
  {{DePoy}}}]{Pepper+2002_SmallTelescopeTransits}%
  \BibitemOpen
  \bibfield  {author} {\bibinfo {author} {\bibfnamefont {J.}~\bibnamefont
  {{Pepper}}}, \bibinfo {author} {\bibfnamefont {A.}~\bibnamefont {{Gould}}}, \
  and\ \bibinfo {author} {\bibfnamefont {D.~L.}\ \bibnamefont {{DePoy}}},\
  }\href@noop {} {\bibfield  {journal} {\bibinfo  {journal} {arXiv e-prints}\
  ,\ \bibinfo {eid} {astro-ph/0209310}} (\bibinfo {year} {2002})},\ \Eprint
  {http://arxiv.org/abs/astro-ph/0209310} {arXiv:astro-ph/0209310 [astro-ph]}
  \BibitemShut {NoStop}%
\bibitem [{\citenamefont {{Bellm}}(2016)}]{Bellm2016_VolumetricRate}%
  \BibitemOpen
  \bibfield  {author} {\bibinfo {author} {\bibfnamefont {E.~C.}\ \bibnamefont
  {{Bellm}}},\ }\href {\doibase 10.1088/1538-3873/128/966/084501} {\bibfield
  {journal} {\bibinfo  {journal} {\pasp}\ }\textbf {\bibinfo {volume} {128}},\
  \bibinfo {pages} {084501} (\bibinfo {year} {2016})},\ \Eprint
  {http://arxiv.org/abs/1605.02081} {arXiv:1605.02081 [astro-ph.IM]}
  \BibitemShut {NoStop}%
\bibitem [{\citenamefont {{Bakos}}\ \emph {et~al.}(2004)\citenamefont
  {{Bakos}}, \citenamefont {{Noyes}}, \citenamefont {{Kov{\'a}cs}},
  \citenamefont {{Stanek}}, \citenamefont {{Sasselov}},\ and\ \citenamefont
  {{Domsa}}}]{Bakos+2004_HAT}%
  \BibitemOpen
  \bibfield  {author} {\bibinfo {author} {\bibfnamefont {G.}~\bibnamefont
  {{Bakos}}}, \bibinfo {author} {\bibfnamefont {R.~W.}\ \bibnamefont
  {{Noyes}}}, \bibinfo {author} {\bibfnamefont {G.}~\bibnamefont
  {{Kov{\'a}cs}}}, \bibinfo {author} {\bibfnamefont {K.~Z.}\ \bibnamefont
  {{Stanek}}}, \bibinfo {author} {\bibfnamefont {D.~D.}\ \bibnamefont
  {{Sasselov}}}, \ and\ \bibinfo {author} {\bibfnamefont {I.}~\bibnamefont
  {{Domsa}}},\ }\href {\doibase 10.1086/382735} {\bibfield  {journal} {\bibinfo
   {journal} {\pasp}\ }\textbf {\bibinfo {volume} {116}},\ \bibinfo {pages}
  {266} (\bibinfo {year} {2004})},\ \Eprint
  {http://arxiv.org/abs/astro-ph/0401219} {arXiv:astro-ph/0401219 [astro-ph]}
  \BibitemShut {NoStop}%
\bibitem [{\citenamefont {{Pollacco}}\ \emph {et~al.}(2006)\citenamefont
  {{Pollacco}}, \citenamefont {{Skillen}}, \citenamefont {{Collier Cameron}},
  \citenamefont {{Christian}}, \citenamefont {{Hellier}}, \citenamefont
  {{Irwin}}, \citenamefont {{Lister}}, \citenamefont {{Street}}, \citenamefont
  {{West}}, \citenamefont {{Anderson}}, \citenamefont {{Clarkson}},
  \citenamefont {{Deeg}}, \citenamefont {{Enoch}}, \citenamefont {{Evans}},
  \citenamefont {{Fitzsimmons}}, \citenamefont {{Haswell}}, \citenamefont
  {{Hodgkin}}, \citenamefont {{Horne}}, \citenamefont {{Kane}}, \citenamefont
  {{Keenan}}, \citenamefont {{Maxted}}, \citenamefont {{Norton}}, \citenamefont
  {{Osborne}}, \citenamefont {{Parley}}, \citenamefont {{Ryans}}, \citenamefont
  {{Smalley}}, \citenamefont {{Wheatley}},\ and\ \citenamefont
  {{Wilson}}}]{Pollacco+2006_WASP_superWASP}%
  \BibitemOpen
  \bibfield  {author} {\bibinfo {author} {\bibfnamefont {D.~L.}\ \bibnamefont
  {{Pollacco}}}, \bibinfo {author} {\bibfnamefont {I.}~\bibnamefont
  {{Skillen}}}, \bibinfo {author} {\bibfnamefont {A.}~\bibnamefont {{Collier
  Cameron}}}, \bibinfo {author} {\bibfnamefont {D.~J.}\ \bibnamefont
  {{Christian}}}, \bibinfo {author} {\bibfnamefont {C.}~\bibnamefont
  {{Hellier}}}, \bibinfo {author} {\bibfnamefont {J.}~\bibnamefont {{Irwin}}},
  \bibinfo {author} {\bibfnamefont {T.~A.}\ \bibnamefont {{Lister}}}, \bibinfo
  {author} {\bibfnamefont {R.~A.}\ \bibnamefont {{Street}}}, \bibinfo {author}
  {\bibfnamefont {R.~G.}\ \bibnamefont {{West}}}, \bibinfo {author}
  {\bibfnamefont {D.~R.}\ \bibnamefont {{Anderson}}}, \bibinfo {author}
  {\bibfnamefont {W.~I.}\ \bibnamefont {{Clarkson}}}, \bibinfo {author}
  {\bibfnamefont {H.}~\bibnamefont {{Deeg}}}, \bibinfo {author} {\bibfnamefont
  {B.}~\bibnamefont {{Enoch}}}, \bibinfo {author} {\bibfnamefont
  {A.}~\bibnamefont {{Evans}}}, \bibinfo {author} {\bibfnamefont
  {A.}~\bibnamefont {{Fitzsimmons}}}, \bibinfo {author} {\bibfnamefont {C.~A.}\
  \bibnamefont {{Haswell}}}, \bibinfo {author} {\bibfnamefont {S.}~\bibnamefont
  {{Hodgkin}}}, \bibinfo {author} {\bibfnamefont {K.}~\bibnamefont {{Horne}}},
  \bibinfo {author} {\bibfnamefont {S.~R.}\ \bibnamefont {{Kane}}}, \bibinfo
  {author} {\bibfnamefont {F.~P.}\ \bibnamefont {{Keenan}}}, \bibinfo {author}
  {\bibfnamefont {P.~F.~L.}\ \bibnamefont {{Maxted}}}, \bibinfo {author}
  {\bibfnamefont {A.~J.}\ \bibnamefont {{Norton}}}, \bibinfo {author}
  {\bibfnamefont {J.}~\bibnamefont {{Osborne}}}, \bibinfo {author}
  {\bibfnamefont {N.~R.}\ \bibnamefont {{Parley}}}, \bibinfo {author}
  {\bibfnamefont {R.~S.~I.}\ \bibnamefont {{Ryans}}}, \bibinfo {author}
  {\bibfnamefont {B.}~\bibnamefont {{Smalley}}}, \bibinfo {author}
  {\bibfnamefont {P.~J.}\ \bibnamefont {{Wheatley}}}, \ and\ \bibinfo {author}
  {\bibfnamefont {D.~M.}\ \bibnamefont {{Wilson}}},\ }\href {\doibase
  10.1086/508556} {\bibfield  {journal} {\bibinfo  {journal} {\pasp}\ }\textbf
  {\bibinfo {volume} {118}},\ \bibinfo {pages} {1407} (\bibinfo {year}
  {2006})},\ \Eprint {http://arxiv.org/abs/astro-ph/0608454}
  {arXiv:astro-ph/0608454 [astro-ph]} \BibitemShut {NoStop}%
\bibitem [{\citenamefont {{Law}}\ \emph {et~al.}(2015)\citenamefont {{Law}},
  \citenamefont {{Fors}}, \citenamefont {{Ratzloff}}, \citenamefont
  {{Wulfken}}, \citenamefont {{Kavanaugh}}, \citenamefont {{Sitar}},
  \citenamefont {{Pruett}}, \citenamefont {{Birchard}}, \citenamefont
  {{Barlow}}, \citenamefont {{Cannon}}, \citenamefont {{Cenko}}, \citenamefont
  {{Dunlap}}, \citenamefont {{Kraus}},\ and\ \citenamefont
  {{Maccarone}}}]{Law+2015_Evryscope_ScienceCase}%
  \BibitemOpen
  \bibfield  {author} {\bibinfo {author} {\bibfnamefont {N.~M.}\ \bibnamefont
  {{Law}}}, \bibinfo {author} {\bibfnamefont {O.}~\bibnamefont {{Fors}}},
  \bibinfo {author} {\bibfnamefont {J.}~\bibnamefont {{Ratzloff}}}, \bibinfo
  {author} {\bibfnamefont {P.}~\bibnamefont {{Wulfken}}}, \bibinfo {author}
  {\bibfnamefont {D.}~\bibnamefont {{Kavanaugh}}}, \bibinfo {author}
  {\bibfnamefont {D.~J.}\ \bibnamefont {{Sitar}}}, \bibinfo {author}
  {\bibfnamefont {Z.}~\bibnamefont {{Pruett}}}, \bibinfo {author}
  {\bibfnamefont {M.~N.}\ \bibnamefont {{Birchard}}}, \bibinfo {author}
  {\bibfnamefont {B.~N.}\ \bibnamefont {{Barlow}}}, \bibinfo {author}
  {\bibfnamefont {K.}~\bibnamefont {{Cannon}}}, \bibinfo {author}
  {\bibfnamefont {S.~B.}\ \bibnamefont {{Cenko}}}, \bibinfo {author}
  {\bibfnamefont {B.}~\bibnamefont {{Dunlap}}}, \bibinfo {author}
  {\bibfnamefont {A.}~\bibnamefont {{Kraus}}}, \ and\ \bibinfo {author}
  {\bibfnamefont {T.~J.}\ \bibnamefont {{Maccarone}}},\ }\href {\doibase
  10.1086/680521} {\bibfield  {journal} {\bibinfo  {journal} {\pasp}\ }\textbf
  {\bibinfo {volume} {127}},\ \bibinfo {pages} {234} (\bibinfo {year}
  {2015})},\ \Eprint {http://arxiv.org/abs/1501.03162} {arXiv:1501.03162
  [astro-ph.IM]} \BibitemShut {NoStop}%
\bibitem [{\citenamefont {{Wheatley}}\ \emph {et~al.}(2018)\citenamefont
  {{Wheatley}}, \citenamefont {{West}}, \citenamefont {{Goad}}, \citenamefont
  {{Jenkins}}, \citenamefont {{Pollacco}}, \citenamefont {{Queloz}},
  \citenamefont {{Rauer}}, \citenamefont {{Udry}}, \citenamefont {{Watson}},
  \citenamefont {{Chazelas}}, \citenamefont {{Eigm{\"u}ller}}, \citenamefont
  {{Lambert}}, \citenamefont {{Genolet}}, \citenamefont {{McCormac}},
  \citenamefont {{Walker}}, \citenamefont {{Armstrong}}, \citenamefont
  {{Bayliss}}, \citenamefont {{Bento}}, \citenamefont {{Bouchy}}, \citenamefont
  {{Burleigh}}, \citenamefont {{Cabrera}}, \citenamefont {{Casewell}},
  \citenamefont {{Chaushev}}, \citenamefont {{Chote}}, \citenamefont
  {{Csizmadia}}, \citenamefont {{Erikson}}, \citenamefont {{Faedi}},
  \citenamefont {{Foxell}}, \citenamefont {{G{\"a}nsicke}}, \citenamefont
  {{Gillen}}, \citenamefont {{Grange}}, \citenamefont {{G{\"u}nther}},
  \citenamefont {{Hodgkin}}, \citenamefont {{Jackman}}, \citenamefont
  {{Jord{\'a}n}}, \citenamefont {{Louden}}, \citenamefont {{Metrailler}},
  \citenamefont {{Moyano}}, \citenamefont {{Nielsen}}, \citenamefont
  {{Osborn}}, \citenamefont {{Poppenhaeger}}, \citenamefont {{Raddi}},
  \citenamefont {{Raynard}}, \citenamefont {{Smith}}, \citenamefont {{Soto}},\
  and\ \citenamefont {{Titz-Weider}}}]{Wheatley+2018_NGTS}%
  \BibitemOpen
  \bibfield  {author} {\bibinfo {author} {\bibfnamefont {P.~J.}\ \bibnamefont
  {{Wheatley}}}, \bibinfo {author} {\bibfnamefont {R.~G.}\ \bibnamefont
  {{West}}}, \bibinfo {author} {\bibfnamefont {M.~R.}\ \bibnamefont {{Goad}}},
  \bibinfo {author} {\bibfnamefont {J.~S.}\ \bibnamefont {{Jenkins}}}, \bibinfo
  {author} {\bibfnamefont {D.~L.}\ \bibnamefont {{Pollacco}}}, \bibinfo
  {author} {\bibfnamefont {D.}~\bibnamefont {{Queloz}}}, \bibinfo {author}
  {\bibfnamefont {H.}~\bibnamefont {{Rauer}}}, \bibinfo {author} {\bibfnamefont
  {S.}~\bibnamefont {{Udry}}}, \bibinfo {author} {\bibfnamefont {C.~A.}\
  \bibnamefont {{Watson}}}, \bibinfo {author} {\bibfnamefont {B.}~\bibnamefont
  {{Chazelas}}}, \bibinfo {author} {\bibfnamefont {P.}~\bibnamefont
  {{Eigm{\"u}ller}}}, \bibinfo {author} {\bibfnamefont {G.}~\bibnamefont
  {{Lambert}}}, \bibinfo {author} {\bibfnamefont {L.}~\bibnamefont
  {{Genolet}}}, \bibinfo {author} {\bibfnamefont {J.}~\bibnamefont
  {{McCormac}}}, \bibinfo {author} {\bibfnamefont {S.}~\bibnamefont
  {{Walker}}}, \bibinfo {author} {\bibfnamefont {D.~J.}\ \bibnamefont
  {{Armstrong}}}, \bibinfo {author} {\bibfnamefont {D.}~\bibnamefont
  {{Bayliss}}}, \bibinfo {author} {\bibfnamefont {J.}~\bibnamefont {{Bento}}},
  \bibinfo {author} {\bibfnamefont {F.}~\bibnamefont {{Bouchy}}}, \bibinfo
  {author} {\bibfnamefont {M.~R.}\ \bibnamefont {{Burleigh}}}, \bibinfo
  {author} {\bibfnamefont {J.}~\bibnamefont {{Cabrera}}}, \bibinfo {author}
  {\bibfnamefont {S.~L.}\ \bibnamefont {{Casewell}}}, \bibinfo {author}
  {\bibfnamefont {A.}~\bibnamefont {{Chaushev}}}, \bibinfo {author}
  {\bibfnamefont {P.}~\bibnamefont {{Chote}}}, \bibinfo {author} {\bibfnamefont
  {S.}~\bibnamefont {{Csizmadia}}}, \bibinfo {author} {\bibfnamefont
  {A.}~\bibnamefont {{Erikson}}}, \bibinfo {author} {\bibfnamefont
  {F.}~\bibnamefont {{Faedi}}}, \bibinfo {author} {\bibfnamefont
  {E.}~\bibnamefont {{Foxell}}}, \bibinfo {author} {\bibfnamefont {B.~T.}\
  \bibnamefont {{G{\"a}nsicke}}}, \bibinfo {author} {\bibfnamefont
  {E.}~\bibnamefont {{Gillen}}}, \bibinfo {author} {\bibfnamefont
  {A.}~\bibnamefont {{Grange}}}, \bibinfo {author} {\bibfnamefont {M.~N.}\
  \bibnamefont {{G{\"u}nther}}}, \bibinfo {author} {\bibfnamefont {S.~T.}\
  \bibnamefont {{Hodgkin}}}, \bibinfo {author} {\bibfnamefont {J.}~\bibnamefont
  {{Jackman}}}, \bibinfo {author} {\bibfnamefont {A.}~\bibnamefont
  {{Jord{\'a}n}}}, \bibinfo {author} {\bibfnamefont {T.}~\bibnamefont
  {{Louden}}}, \bibinfo {author} {\bibfnamefont {L.}~\bibnamefont
  {{Metrailler}}}, \bibinfo {author} {\bibfnamefont {M.}~\bibnamefont
  {{Moyano}}}, \bibinfo {author} {\bibfnamefont {L.~D.}\ \bibnamefont
  {{Nielsen}}}, \bibinfo {author} {\bibfnamefont {H.~P.}\ \bibnamefont
  {{Osborn}}}, \bibinfo {author} {\bibfnamefont {K.}~\bibnamefont
  {{Poppenhaeger}}}, \bibinfo {author} {\bibfnamefont {R.}~\bibnamefont
  {{Raddi}}}, \bibinfo {author} {\bibfnamefont {L.}~\bibnamefont {{Raynard}}},
  \bibinfo {author} {\bibfnamefont {A.~M.~S.}\ \bibnamefont {{Smith}}},
  \bibinfo {author} {\bibfnamefont {M.}~\bibnamefont {{Soto}}}, \ and\ \bibinfo
  {author} {\bibfnamefont {R.}~\bibnamefont {{Titz-Weider}}},\ }\href {\doibase
  10.1093/mnras/stx2836} {\bibfield  {journal} {\bibinfo  {journal} {\mnras}\
  }\textbf {\bibinfo {volume} {475}},\ \bibinfo {pages} {4476} (\bibinfo {year}
  {2018})},\ \Eprint {http://arxiv.org/abs/1710.11100} {arXiv:1710.11100
  [astro-ph.EP]} \BibitemShut {NoStop}%
\bibitem [{\citenamefont {{Heinze}}\ \emph {et~al.}(2018)\citenamefont
  {{Heinze}}, \citenamefont {{Tonry}}, \citenamefont {{Denneau}}, \citenamefont
  {{Flewelling}}, \citenamefont {{Stalder}}, \citenamefont {{Rest}},
  \citenamefont {{Smith}}, \citenamefont {{Smartt}},\ and\ \citenamefont
  {{Weiland}}}]{Heinze+2018_ATLAS_VarStars}%
  \BibitemOpen
  \bibfield  {author} {\bibinfo {author} {\bibfnamefont {A.~N.}\ \bibnamefont
  {{Heinze}}}, \bibinfo {author} {\bibfnamefont {J.~L.}\ \bibnamefont
  {{Tonry}}}, \bibinfo {author} {\bibfnamefont {L.}~\bibnamefont {{Denneau}}},
  \bibinfo {author} {\bibfnamefont {H.}~\bibnamefont {{Flewelling}}}, \bibinfo
  {author} {\bibfnamefont {B.}~\bibnamefont {{Stalder}}}, \bibinfo {author}
  {\bibfnamefont {A.}~\bibnamefont {{Rest}}}, \bibinfo {author} {\bibfnamefont
  {K.~W.}\ \bibnamefont {{Smith}}}, \bibinfo {author} {\bibfnamefont {S.~J.}\
  \bibnamefont {{Smartt}}}, \ and\ \bibinfo {author} {\bibfnamefont
  {H.}~\bibnamefont {{Weiland}}},\ }\href {\doibase 10.3847/1538-3881/aae47f}
  {\bibfield  {journal} {\bibinfo  {journal} {\aj}\ }\textbf {\bibinfo {volume}
  {156}},\ \bibinfo {eid} {241} (\bibinfo {year} {2018})},\ \Eprint
  {http://arxiv.org/abs/1804.02132} {arXiv:1804.02132 [astro-ph.SR]}
  \BibitemShut {NoStop}%
\bibitem [{\citenamefont {{Abraham}}\ and\ \citenamefont {{van
  Dokkum}}(2014)}]{Abraham+2014_DragonFly}%
  \BibitemOpen
  \bibfield  {author} {\bibinfo {author} {\bibfnamefont {R.~G.}\ \bibnamefont
  {{Abraham}}}\ and\ \bibinfo {author} {\bibfnamefont {P.~G.}\ \bibnamefont
  {{van Dokkum}}},\ }\href {\doibase 10.1086/674875} {\bibfield  {journal}
  {\bibinfo  {journal} {\pasp}\ }\textbf {\bibinfo {volume} {126}},\ \bibinfo
  {pages} {55} (\bibinfo {year} {2014})},\ \Eprint
  {http://arxiv.org/abs/1401.5473} {arXiv:1401.5473 [astro-ph.IM]} \BibitemShut
  {NoStop}%
\bibitem [{\citenamefont {{Danieli}}\ \emph {et~al.}(2018)\citenamefont
  {{Danieli}}, \citenamefont {{van Dokkum}},\ and\ \citenamefont
  {{Conroy}}}]{Danieli+2018_DwarfGalaxiesInTheField_DragonFly}%
  \BibitemOpen
  \bibfield  {author} {\bibinfo {author} {\bibfnamefont {S.}~\bibnamefont
  {{Danieli}}}, \bibinfo {author} {\bibfnamefont {P.}~\bibnamefont {{van
  Dokkum}}}, \ and\ \bibinfo {author} {\bibfnamefont {C.}~\bibnamefont
  {{Conroy}}},\ }\href {\doibase 10.3847/1538-4357/aaadfb} {\bibfield
  {journal} {\bibinfo  {journal} {\apj}\ }\textbf {\bibinfo {volume} {856}},\
  \bibinfo {eid} {69} (\bibinfo {year} {2018})},\ \Eprint
  {http://arxiv.org/abs/1711.00860} {arXiv:1711.00860 [astro-ph.GA]}
  \BibitemShut {NoStop}%
\bibitem [{\citenamefont {{Bellm}}\ \emph
  {et~al.}(2019{\natexlab{b}})\citenamefont {{Bellm}}, \citenamefont
  {{Kulkarni}}, \citenamefont {{Barlow}}, \citenamefont {{Feindt}},
  \citenamefont {{Graham}}, \citenamefont {{Goobar}}, \citenamefont {{Kupfer}},
  \citenamefont {{Ngeow}}, \citenamefont {{Nugent}},\ and\ \citenamefont
  {{Ofek}}}]{Bellm+2019_ZTF_Scheduler}%
  \BibitemOpen
  \bibfield  {author} {\bibinfo {author} {\bibfnamefont {E.~C.}\ \bibnamefont
  {{Bellm}}}, \bibinfo {author} {\bibfnamefont {S.~R.}\ \bibnamefont
  {{Kulkarni}}}, \bibinfo {author} {\bibfnamefont {T.}~\bibnamefont
  {{Barlow}}}, \bibinfo {author} {\bibfnamefont {U.}~\bibnamefont {{Feindt}}},
  \bibinfo {author} {\bibfnamefont {M.~J.}\ \bibnamefont {{Graham}}}, \bibinfo
  {author} {\bibfnamefont {A.}~\bibnamefont {{Goobar}}}, \bibinfo {author}
  {\bibfnamefont {T.}~\bibnamefont {{Kupfer}}}, \bibinfo {author}
  {\bibfnamefont {C.-C.}\ \bibnamefont {{Ngeow}}}, \bibinfo {author}
  {\bibfnamefont {P.}~\bibnamefont {{Nugent}}}, \ and\ \bibinfo {author}
  {\bibfnamefont {E.}~\bibnamefont {{Ofek}}},\ }\href {\doibase
  10.1088/1538-3873/ab0c2a} {\bibfield  {journal} {\bibinfo  {journal} {\pasp}\
  }\textbf {\bibinfo {volume} {131}},\ \bibinfo {pages} {068003} (\bibinfo
  {year} {2019}{\natexlab{b}})},\ \Eprint {http://arxiv.org/abs/1905.02209}
  {arXiv:1905.02209 [astro-ph.IM]} \BibitemShut {NoStop}%
\bibitem [{\citenamefont {{Zackay}}\ and\ \citenamefont
  {{Ofek}}(2017{\natexlab{a}})}]{Zackay+2017_CoadditionI}%
  \BibitemOpen
  \bibfield  {author} {\bibinfo {author} {\bibfnamefont {B.}~\bibnamefont
  {{Zackay}}}\ and\ \bibinfo {author} {\bibfnamefont {E.~O.}\ \bibnamefont
  {{Ofek}}},\ }\href {\doibase 10.3847/1538-4357/836/2/187} {\bibfield
  {journal} {\bibinfo  {journal} {\apj}\ }\textbf {\bibinfo {volume} {836}},\
  \bibinfo {eid} {187} (\bibinfo {year} {2017}{\natexlab{a}})},\ \Eprint
  {http://arxiv.org/abs/1512.06872} {arXiv:1512.06872 [astro-ph.IM]}
  \BibitemShut {NoStop}%
\bibitem [{\citenamefont {{Nir}}\ \emph {et~al.}(2019)\citenamefont {{Nir}},
  \citenamefont {{Zackay}},\ and\ \citenamefont
  {{Ofek}}}]{Nir+2019_NoncircularPupilTelescope}%
  \BibitemOpen
  \bibfield  {author} {\bibinfo {author} {\bibfnamefont {G.}~\bibnamefont
  {{Nir}}}, \bibinfo {author} {\bibfnamefont {B.}~\bibnamefont {{Zackay}}}, \
  and\ \bibinfo {author} {\bibfnamefont {E.~O.}\ \bibnamefont {{Ofek}}},\
  }\href {\doibase 10.3847/1538-3881/ab27c7} {\bibfield  {journal} {\bibinfo
  {journal} {\aj}\ }\textbf {\bibinfo {volume} {158}},\ \bibinfo {eid} {70}
  (\bibinfo {year} {2019})},\ \Eprint {http://arxiv.org/abs/1809.09933}
  {arXiv:1809.09933 [astro-ph.IM]} \BibitemShut {NoStop}%
\bibitem [{\citenamefont
  {{Labeyrie}}(1970)}]{Labeyrie1970_SpeckleInterferometry}%
  \BibitemOpen
  \bibfield  {author} {\bibinfo {author} {\bibfnamefont {A.}~\bibnamefont
  {{Labeyrie}}},\ }\href@noop {} {\bibfield  {journal} {\bibinfo  {journal}
  {\aap}\ }\textbf {\bibinfo {volume} {6}},\ \bibinfo {pages} {85} (\bibinfo
  {year} {1970})}\BibitemShut {NoStop}%
\bibitem [{\citenamefont
  {{Bates}}(1982)}]{Bates1982_SpeckleInterferometryReview}%
  \BibitemOpen
  \bibfield  {author} {\bibinfo {author} {\bibfnamefont {R.~H.~T.}\
  \bibnamefont {{Bates}}},\ }\href {\doibase 10.1016/0370-1573(82)90121-1}
  {\bibfield  {journal} {\bibinfo  {journal} {\physrep}\ }\textbf {\bibinfo
  {volume} {90}},\ \bibinfo {pages} {203} (\bibinfo {year} {1982})}\BibitemShut
  {NoStop}%
\bibitem [{\citenamefont {{Zackay}}\ and\ \citenamefont
  {{Ofek}}(2017{\natexlab{b}})}]{Zackay+2017_CoadditionII}%
  \BibitemOpen
  \bibfield  {author} {\bibinfo {author} {\bibfnamefont {B.}~\bibnamefont
  {{Zackay}}}\ and\ \bibinfo {author} {\bibfnamefont {E.~O.}\ \bibnamefont
  {{Ofek}}},\ }\href {\doibase 10.3847/1538-4357/836/2/188} {\bibfield
  {journal} {\bibinfo  {journal} {\apj}\ }\textbf {\bibinfo {volume} {836}},\
  \bibinfo {eid} {188} (\bibinfo {year} {2017}{\natexlab{b}})},\ \Eprint
  {http://arxiv.org/abs/1512.06879} {arXiv:1512.06879 [astro-ph.IM]}
  \BibitemShut {NoStop}%
\bibitem [{\citenamefont {{Ratzloff}}\ \emph {et~al.}(2019)\citenamefont
  {{Ratzloff}}, \citenamefont {{Law}}, \citenamefont {{Fors}}, \citenamefont
  {{Corbett}}, \citenamefont {{Howard}}, \citenamefont {{del Ser}},\ and\
  \citenamefont {{Haislip}}}]{Ratzloff+2019_EvryScope}%
  \BibitemOpen
  \bibfield  {author} {\bibinfo {author} {\bibfnamefont {J.~K.}\ \bibnamefont
  {{Ratzloff}}}, \bibinfo {author} {\bibfnamefont {N.~M.}\ \bibnamefont
  {{Law}}}, \bibinfo {author} {\bibfnamefont {O.}~\bibnamefont {{Fors}}},
  \bibinfo {author} {\bibfnamefont {H.~T.}\ \bibnamefont {{Corbett}}}, \bibinfo
  {author} {\bibfnamefont {W.~S.}\ \bibnamefont {{Howard}}}, \bibinfo {author}
  {\bibfnamefont {D.}~\bibnamefont {{del Ser}}}, \ and\ \bibinfo {author}
  {\bibfnamefont {J.}~\bibnamefont {{Haislip}}},\ }\href {\doibase
  10.1088/1538-3873/ab19d0} {\bibfield  {journal} {\bibinfo  {journal} {\pasp}\
  }\textbf {\bibinfo {volume} {131}},\ \bibinfo {pages} {075001} (\bibinfo
  {year} {2019})},\ \Eprint {http://arxiv.org/abs/1904.11991} {arXiv:1904.11991
  [astro-ph.IM]} \BibitemShut {NoStop}%
\bibitem [{\citenamefont {{Cucciati}}\ \emph {et~al.}(2012)\citenamefont
  {{Cucciati}}, \citenamefont {{Tresse}}, \citenamefont {{Ilbert}},
  \citenamefont {{Le F{\`e}vre}}, \citenamefont {{Garilli}}, \citenamefont {{Le
  Brun}}, \citenamefont {{Cassata}}, \citenamefont {{Franzetti}}, \citenamefont
  {{Maccagni}}, \citenamefont {{Scodeggio}}, \citenamefont {{Zucca}},
  \citenamefont {{Zamorani}}, \citenamefont {{Bardelli}}, \citenamefont
  {{Bolzonella}}, \citenamefont {{Bielby}}, \citenamefont {{McCracken}},
  \citenamefont {{Zanichelli}},\ and\ \citenamefont
  {{Vergani}}}]{Cucciati+2012_SFRz}%
  \BibitemOpen
  \bibfield  {author} {\bibinfo {author} {\bibfnamefont {O.}~\bibnamefont
  {{Cucciati}}}, \bibinfo {author} {\bibfnamefont {L.}~\bibnamefont
  {{Tresse}}}, \bibinfo {author} {\bibfnamefont {O.}~\bibnamefont {{Ilbert}}},
  \bibinfo {author} {\bibfnamefont {O.}~\bibnamefont {{Le F{\`e}vre}}},
  \bibinfo {author} {\bibfnamefont {B.}~\bibnamefont {{Garilli}}}, \bibinfo
  {author} {\bibfnamefont {V.}~\bibnamefont {{Le Brun}}}, \bibinfo {author}
  {\bibfnamefont {P.}~\bibnamefont {{Cassata}}}, \bibinfo {author}
  {\bibfnamefont {P.}~\bibnamefont {{Franzetti}}}, \bibinfo {author}
  {\bibfnamefont {D.}~\bibnamefont {{Maccagni}}}, \bibinfo {author}
  {\bibfnamefont {M.}~\bibnamefont {{Scodeggio}}}, \bibinfo {author}
  {\bibfnamefont {E.}~\bibnamefont {{Zucca}}}, \bibinfo {author} {\bibfnamefont
  {G.}~\bibnamefont {{Zamorani}}}, \bibinfo {author} {\bibfnamefont
  {S.}~\bibnamefont {{Bardelli}}}, \bibinfo {author} {\bibfnamefont
  {M.}~\bibnamefont {{Bolzonella}}}, \bibinfo {author} {\bibfnamefont {R.~M.}\
  \bibnamefont {{Bielby}}}, \bibinfo {author} {\bibfnamefont {H.~J.}\
  \bibnamefont {{McCracken}}}, \bibinfo {author} {\bibfnamefont
  {A.}~\bibnamefont {{Zanichelli}}}, \ and\ \bibinfo {author} {\bibfnamefont
  {D.}~\bibnamefont {{Vergani}}},\ }\href {\doibase
  10.1051/0004-6361/201118010} {\bibfield  {journal} {\bibinfo  {journal}
  {\aap}\ }\textbf {\bibinfo {volume} {539}},\ \bibinfo {eid} {A31} (\bibinfo
  {year} {2012})},\ \Eprint {http://arxiv.org/abs/1109.1005} {arXiv:1109.1005
  [astro-ph.CO]} \BibitemShut {NoStop}%
\bibitem [{\citenamefont {{Planck Collaboration}}\ \emph
  {et~al.}(2016)\citenamefont {{Planck Collaboration}}, \citenamefont {{Ade}},
  \citenamefont {{Aghanim}}, \citenamefont {{Arnaud}}, \citenamefont
  {{Ashdown}}, \citenamefont {{Aumont}}, \citenamefont {{Baccigalupi}},
  \citenamefont {{Banday}}, \citenamefont {{Barreiro}}, \citenamefont
  {{Bartlett}}, \citenamefont {{Bartolo}}, \citenamefont {{Battaner}},
  \citenamefont {{Battye}}, \citenamefont {{Benabed}}, \citenamefont
  {{Beno{\^\i}t}}, \citenamefont {{Benoit-L{\'e}vy}}, \citenamefont
  {{Bernard}}, \citenamefont {{Bersanelli}}, \citenamefont {{Bielewicz}},
  \citenamefont {{Bock}}, \citenamefont {{Bonaldi}}, \citenamefont
  {{Bonavera}}, \citenamefont {{Bond}}, \citenamefont {{Borrill}},
  \citenamefont {{Bouchet}}, \citenamefont {{Boulanger}}, \citenamefont
  {{Bucher}}, \citenamefont {{Burigana}}, \citenamefont {{Butler}},
  \citenamefont {{Calabrese}}, \citenamefont {{Cardoso}}, \citenamefont
  {{Catalano}}, \citenamefont {{Challinor}}, \citenamefont {{Chamballu}},
  \citenamefont {{Chary}}, \citenamefont {{Chiang}}, \citenamefont {{Chluba}},
  \citenamefont {{Christensen}}, \citenamefont {{Church}}, \citenamefont
  {{Clements}}, \citenamefont {{Colombi}}, \citenamefont {{Colombo}},
  \citenamefont {{Combet}}, \citenamefont {{Coulais}}, \citenamefont {{Crill}},
  \citenamefont {{Curto}}, \citenamefont {{Cuttaia}}, \citenamefont {{Danese}},
  \citenamefont {{Davies}}, \citenamefont {{Davis}}, \citenamefont {{de
  Bernardis}}, \citenamefont {{de Rosa}}, \citenamefont {{de Zotti}},
  \citenamefont {{Delabrouille}}, \citenamefont {{D{\'e}sert}}, \citenamefont
  {{Di Valentino}}, \citenamefont {{Dickinson}}, \citenamefont {{Diego}},
  \citenamefont {{Dolag}}, \citenamefont {{Dole}}, \citenamefont {{Donzelli}},
  \citenamefont {{Dor{\'e}}}, \citenamefont {{Douspis}}, \citenamefont
  {{Ducout}}, \citenamefont {{Dunkley}}, \citenamefont {{Dupac}}, \citenamefont
  {{Efstathiou}}, \citenamefont {{Elsner}}, \citenamefont {{En{\ss}lin}},
  \citenamefont {{Eriksen}}, \citenamefont {{Farhang}}, \citenamefont
  {{Fergusson}}, \citenamefont {{Finelli}}, \citenamefont {{Forni}},
  \citenamefont {{Frailis}}, \citenamefont {{Fraisse}}, \citenamefont
  {{Franceschi}}, \citenamefont {{Frejsel}}, \citenamefont {{Galeotta}},
  \citenamefont {{Galli}}, \citenamefont {{Ganga}}, \citenamefont {{Gauthier}},
  \citenamefont {{Gerbino}}, \citenamefont {{Ghosh}}, \citenamefont {{Giard}},
  \citenamefont {{Giraud-H{\'e}raud}}, \citenamefont {{Giusarma}},
  \citenamefont {{Gjerl{\o}w}}, \citenamefont {{Gonz{\'a}lez-Nuevo}},
  \citenamefont {{G{\'o}rski}}, \citenamefont {{Gratton}}, \citenamefont
  {{Gregorio}}, \citenamefont {{Gruppuso}}, \citenamefont {{Gudmundsson}},
  \citenamefont {{Hamann}}, \citenamefont {{Hansen}}, \citenamefont {{Hanson}},
  \citenamefont {{Harrison}}, \citenamefont {{Helou}}, \citenamefont
  {{Henrot-Versill{\'e}}}, \citenamefont {{Hern{\'a}ndez-Monteagudo}},
  \citenamefont {{Herranz}}, \citenamefont {{Hildebrand t}}, \citenamefont
  {{Hivon}}, \citenamefont {{Hobson}}, \citenamefont {{Holmes}}, \citenamefont
  {{Hornstrup}}, \citenamefont {{Hovest}}, \citenamefont {{Huang}},
  \citenamefont {{Huffenberger}}, \citenamefont {{Hurier}}, \citenamefont
  {{Jaffe}}, \citenamefont {{Jaffe}}, \citenamefont {{Jones}}, \citenamefont
  {{Juvela}}, \citenamefont {{Keih{\"a}nen}}, \citenamefont {{Keskitalo}},
  \citenamefont {{Kisner}}, \citenamefont {{Kneissl}}, \citenamefont
  {{Knoche}}, \citenamefont {{Knox}}, \citenamefont {{Kunz}}, \citenamefont
  {{Kurki-Suonio}}, \citenamefont {{Lagache}}, \citenamefont
  {{L{\"a}hteenm{\"a}ki}}, \citenamefont {{Lamarre}}, \citenamefont
  {{Lasenby}}, \citenamefont {{Lattanzi}}, \citenamefont {{Lawrence}},
  \citenamefont {{Leahy}}, \citenamefont {{Leonardi}}, \citenamefont
  {{Lesgourgues}}, \citenamefont {{Levrier}}, \citenamefont {{Lewis}},
  \citenamefont {{Liguori}}, \citenamefont {{Lilje}}, \citenamefont
  {{Linden-V{\o}rnle}}, \citenamefont {{L{\'o}pez-Caniego}}, \citenamefont
  {{Lubin}}, \citenamefont {{Mac{\'\i}as-P{\'e}rez}}, \citenamefont {{Maggio}},
  \citenamefont {{Maino}}, \citenamefont {{Mandolesi}}, \citenamefont
  {{Mangilli}}, \citenamefont {{Marchini}}, \citenamefont {{Maris}},
  \citenamefont {{Martin}}, \citenamefont {{Martinelli}}, \citenamefont
  {{Mart{\'\i}nez-Gonz{\'a}lez}}, \citenamefont {{Masi}}, \citenamefont
  {{Matarrese}}, \citenamefont {{McGehee}}, \citenamefont {{Meinhold}},
  \citenamefont {{Melchiorri}}, \citenamefont {{Melin}}, \citenamefont
  {{Mendes}}, \citenamefont {{Mennella}}, \citenamefont {{Migliaccio}},
  \citenamefont {{Millea}}, \citenamefont {{Mitra}}, \citenamefont
  {{Miville-Desch{\^e}nes}}, \citenamefont {{Moneti}}, \citenamefont
  {{Montier}}, \citenamefont {{Morgante}}, \citenamefont {{Mortlock}},
  \citenamefont {{Moss}}, \citenamefont {{Munshi}}, \citenamefont {{Murphy}},
  \citenamefont {{Naselsky}}, \citenamefont {{Nati}}, \citenamefont {{Natoli}},
  \citenamefont {{Netterfield}}, \citenamefont {{N{\o}rgaard-Nielsen}},
  \citenamefont {{Noviello}}, \citenamefont {{Novikov}}, \citenamefont
  {{Novikov}}, \citenamefont {{Oxborrow}}, \citenamefont {{Paci}},
  \citenamefont {{Pagano}}, \citenamefont {{Pajot}}, \citenamefont
  {{Paladini}}, \citenamefont {{Paoletti}}, \citenamefont {{Partridge}},
  \citenamefont {{Pasian}}, \citenamefont {{Patanchon}}, \citenamefont
  {{Pearson}}, \citenamefont {{Perdereau}}, \citenamefont {{Perotto}},
  \citenamefont {{Perrotta}}, \citenamefont {{Pettorino}}, \citenamefont
  {{Piacentini}}, \citenamefont {{Piat}}, \citenamefont {{Pierpaoli}},
  \citenamefont {{Pietrobon}}, \citenamefont {{Plaszczynski}}, \citenamefont
  {{Pointecouteau}}, \citenamefont {{Polenta}}, \citenamefont {{Popa}},
  \citenamefont {{Pratt}}, \citenamefont {{Pr{\'e}zeau}}, \citenamefont
  {{Prunet}}, \citenamefont {{Puget}}, \citenamefont {{Rachen}}, \citenamefont
  {{Reach}}, \citenamefont {{Rebolo}}, \citenamefont {{Reinecke}},
  \citenamefont {{Remazeilles}}, \citenamefont {{Renault}}, \citenamefont
  {{Renzi}}, \citenamefont {{Ristorcelli}}, \citenamefont {{Rocha}},
  \citenamefont {{Rosset}}, \citenamefont {{Rossetti}}, \citenamefont
  {{Roudier}}, \citenamefont {{Rouill{\'e} d'Orfeuil}}, \citenamefont
  {{Rowan-Robinson}}, \citenamefont {{Rubi{\~n}o-Mart{\'\i}n}}, \citenamefont
  {{Rusholme}}, \citenamefont {{Said}}, \citenamefont {{Salvatelli}},
  \citenamefont {{Salvati}}, \citenamefont {{Sandri}}, \citenamefont
  {{Santos}}, \citenamefont {{Savelainen}}, \citenamefont {{Savini}},
  \citenamefont {{Scott}}, \citenamefont {{Seiffert}}, \citenamefont {{Serra}},
  \citenamefont {{Shellard}}, \citenamefont {{Spencer}}, \citenamefont
  {{Spinelli}}, \citenamefont {{Stolyarov}}, \citenamefont {{Stompor}},
  \citenamefont {{Sudiwala}}, \citenamefont {{Sunyaev}}, \citenamefont
  {{Sutton}}, \citenamefont {{Suur-Uski}}, \citenamefont {{Sygnet}},
  \citenamefont {{Tauber}}, \citenamefont {{Terenzi}}, \citenamefont
  {{Toffolatti}}, \citenamefont {{Tomasi}}, \citenamefont {{Tristram}},
  \citenamefont {{Trombetti}}, \citenamefont {{Tucci}}, \citenamefont
  {{Tuovinen}}, \citenamefont {{T{\"u}rler}}, \citenamefont {{Umana}},
  \citenamefont {{Valenziano}}, \citenamefont {{Valiviita}}, \citenamefont
  {{Van Tent}}, \citenamefont {{Vielva}}, \citenamefont {{Villa}},
  \citenamefont {{Wade}}, \citenamefont {{Wandelt}}, \citenamefont {{Wehus}},
  \citenamefont {{White}}, \citenamefont {{White}}, \citenamefont
  {{Wilkinson}}, \citenamefont {{Yvon}}, \citenamefont {{Zacchei}},\ and\
  \citenamefont {{Zonca}}}]{Ade+2016_Planck2015_CosmologicalParameters}%
  \BibitemOpen
  \bibfield  {author} {\bibinfo {author} {\bibnamefont {{Planck
  Collaboration}}}, \bibinfo {author} {\bibfnamefont {P.~A.~R.}\ \bibnamefont
  {{Ade}}}, \bibinfo {author} {\bibfnamefont {N.}~\bibnamefont {{Aghanim}}},
  \bibinfo {author} {\bibfnamefont {M.}~\bibnamefont {{Arnaud}}}, \bibinfo
  {author} {\bibfnamefont {M.}~\bibnamefont {{Ashdown}}}, \bibinfo {author}
  {\bibfnamefont {J.}~\bibnamefont {{Aumont}}}, \bibinfo {author}
  {\bibfnamefont {C.}~\bibnamefont {{Baccigalupi}}}, \bibinfo {author}
  {\bibfnamefont {A.~J.}\ \bibnamefont {{Banday}}}, \bibinfo {author}
  {\bibfnamefont {R.~B.}\ \bibnamefont {{Barreiro}}}, \bibinfo {author}
  {\bibfnamefont {J.~G.}\ \bibnamefont {{Bartlett}}}, \bibinfo {author}
  {\bibfnamefont {N.}~\bibnamefont {{Bartolo}}}, \bibinfo {author}
  {\bibfnamefont {E.}~\bibnamefont {{Battaner}}}, \bibinfo {author}
  {\bibfnamefont {R.}~\bibnamefont {{Battye}}}, \bibinfo {author}
  {\bibfnamefont {K.}~\bibnamefont {{Benabed}}}, \bibinfo {author}
  {\bibfnamefont {A.}~\bibnamefont {{Beno{\^\i}t}}}, \bibinfo {author}
  {\bibfnamefont {A.}~\bibnamefont {{Benoit-L{\'e}vy}}}, \bibinfo {author}
  {\bibfnamefont {J.~P.}\ \bibnamefont {{Bernard}}}, \bibinfo {author}
  {\bibfnamefont {M.}~\bibnamefont {{Bersanelli}}}, \bibinfo {author}
  {\bibfnamefont {P.}~\bibnamefont {{Bielewicz}}}, \bibinfo {author}
  {\bibfnamefont {J.~J.}\ \bibnamefont {{Bock}}}, \bibinfo {author}
  {\bibfnamefont {A.}~\bibnamefont {{Bonaldi}}}, \bibinfo {author}
  {\bibfnamefont {L.}~\bibnamefont {{Bonavera}}}, \bibinfo {author}
  {\bibfnamefont {J.~R.}\ \bibnamefont {{Bond}}}, \bibinfo {author}
  {\bibfnamefont {J.}~\bibnamefont {{Borrill}}}, \bibinfo {author}
  {\bibfnamefont {F.~R.}\ \bibnamefont {{Bouchet}}}, \bibinfo {author}
  {\bibfnamefont {F.}~\bibnamefont {{Boulanger}}}, \bibinfo {author}
  {\bibfnamefont {M.}~\bibnamefont {{Bucher}}}, \bibinfo {author}
  {\bibfnamefont {C.}~\bibnamefont {{Burigana}}}, \bibinfo {author}
  {\bibfnamefont {R.~C.}\ \bibnamefont {{Butler}}}, \bibinfo {author}
  {\bibfnamefont {E.}~\bibnamefont {{Calabrese}}}, \bibinfo {author}
  {\bibfnamefont {J.~F.}\ \bibnamefont {{Cardoso}}}, \bibinfo {author}
  {\bibfnamefont {A.}~\bibnamefont {{Catalano}}}, \bibinfo {author}
  {\bibfnamefont {A.}~\bibnamefont {{Challinor}}}, \bibinfo {author}
  {\bibfnamefont {A.}~\bibnamefont {{Chamballu}}}, \bibinfo {author}
  {\bibfnamefont {R.~R.}\ \bibnamefont {{Chary}}}, \bibinfo {author}
  {\bibfnamefont {H.~C.}\ \bibnamefont {{Chiang}}}, \bibinfo {author}
  {\bibfnamefont {J.}~\bibnamefont {{Chluba}}}, \bibinfo {author}
  {\bibfnamefont {P.~R.}\ \bibnamefont {{Christensen}}}, \bibinfo {author}
  {\bibfnamefont {S.}~\bibnamefont {{Church}}}, \bibinfo {author}
  {\bibfnamefont {D.~L.}\ \bibnamefont {{Clements}}}, \bibinfo {author}
  {\bibfnamefont {S.}~\bibnamefont {{Colombi}}}, \bibinfo {author}
  {\bibfnamefont {L.~P.~L.}\ \bibnamefont {{Colombo}}}, \bibinfo {author}
  {\bibfnamefont {C.}~\bibnamefont {{Combet}}}, \bibinfo {author}
  {\bibfnamefont {A.}~\bibnamefont {{Coulais}}}, \bibinfo {author}
  {\bibfnamefont {B.~P.}\ \bibnamefont {{Crill}}}, \bibinfo {author}
  {\bibfnamefont {A.}~\bibnamefont {{Curto}}}, \bibinfo {author} {\bibfnamefont
  {F.}~\bibnamefont {{Cuttaia}}}, \bibinfo {author} {\bibfnamefont
  {L.}~\bibnamefont {{Danese}}}, \bibinfo {author} {\bibfnamefont {R.~D.}\
  \bibnamefont {{Davies}}}, \bibinfo {author} {\bibfnamefont {R.~J.}\
  \bibnamefont {{Davis}}}, \bibinfo {author} {\bibfnamefont {P.}~\bibnamefont
  {{de Bernardis}}}, \bibinfo {author} {\bibfnamefont {A.}~\bibnamefont {{de
  Rosa}}}, \bibinfo {author} {\bibfnamefont {G.}~\bibnamefont {{de Zotti}}},
  \bibinfo {author} {\bibfnamefont {J.}~\bibnamefont {{Delabrouille}}},
  \bibinfo {author} {\bibfnamefont {F.~X.}\ \bibnamefont {{D{\'e}sert}}},
  \bibinfo {author} {\bibfnamefont {E.}~\bibnamefont {{Di Valentino}}},
  \bibinfo {author} {\bibfnamefont {C.}~\bibnamefont {{Dickinson}}}, \bibinfo
  {author} {\bibfnamefont {J.~M.}\ \bibnamefont {{Diego}}}, \bibinfo {author}
  {\bibfnamefont {K.}~\bibnamefont {{Dolag}}}, \bibinfo {author} {\bibfnamefont
  {H.}~\bibnamefont {{Dole}}}, \bibinfo {author} {\bibfnamefont
  {S.}~\bibnamefont {{Donzelli}}}, \bibinfo {author} {\bibfnamefont
  {O.}~\bibnamefont {{Dor{\'e}}}}, \bibinfo {author} {\bibfnamefont
  {M.}~\bibnamefont {{Douspis}}}, \bibinfo {author} {\bibfnamefont
  {A.}~\bibnamefont {{Ducout}}}, \bibinfo {author} {\bibfnamefont
  {J.}~\bibnamefont {{Dunkley}}}, \bibinfo {author} {\bibfnamefont
  {X.}~\bibnamefont {{Dupac}}}, \bibinfo {author} {\bibfnamefont
  {G.}~\bibnamefont {{Efstathiou}}}, \bibinfo {author} {\bibfnamefont
  {F.}~\bibnamefont {{Elsner}}}, \bibinfo {author} {\bibfnamefont {T.~A.}\
  \bibnamefont {{En{\ss}lin}}}, \bibinfo {author} {\bibfnamefont {H.~K.}\
  \bibnamefont {{Eriksen}}}, \bibinfo {author} {\bibfnamefont {M.}~\bibnamefont
  {{Farhang}}}, \bibinfo {author} {\bibfnamefont {J.}~\bibnamefont
  {{Fergusson}}}, \bibinfo {author} {\bibfnamefont {F.}~\bibnamefont
  {{Finelli}}}, \bibinfo {author} {\bibfnamefont {O.}~\bibnamefont {{Forni}}},
  \bibinfo {author} {\bibfnamefont {M.}~\bibnamefont {{Frailis}}}, \bibinfo
  {author} {\bibfnamefont {A.~A.}\ \bibnamefont {{Fraisse}}}, \bibinfo {author}
  {\bibfnamefont {E.}~\bibnamefont {{Franceschi}}}, \bibinfo {author}
  {\bibfnamefont {A.}~\bibnamefont {{Frejsel}}}, \bibinfo {author}
  {\bibfnamefont {S.}~\bibnamefont {{Galeotta}}}, \bibinfo {author}
  {\bibfnamefont {S.}~\bibnamefont {{Galli}}}, \bibinfo {author} {\bibfnamefont
  {K.}~\bibnamefont {{Ganga}}}, \bibinfo {author} {\bibfnamefont
  {C.}~\bibnamefont {{Gauthier}}}, \bibinfo {author} {\bibfnamefont
  {M.}~\bibnamefont {{Gerbino}}}, \bibinfo {author} {\bibfnamefont
  {T.}~\bibnamefont {{Ghosh}}}, \bibinfo {author} {\bibfnamefont
  {M.}~\bibnamefont {{Giard}}}, \bibinfo {author} {\bibfnamefont
  {Y.}~\bibnamefont {{Giraud-H{\'e}raud}}}, \bibinfo {author} {\bibfnamefont
  {E.}~\bibnamefont {{Giusarma}}}, \bibinfo {author} {\bibfnamefont
  {E.}~\bibnamefont {{Gjerl{\o}w}}}, \bibinfo {author} {\bibfnamefont
  {J.}~\bibnamefont {{Gonz{\'a}lez-Nuevo}}}, \bibinfo {author} {\bibfnamefont
  {K.~M.}\ \bibnamefont {{G{\'o}rski}}}, \bibinfo {author} {\bibfnamefont
  {S.}~\bibnamefont {{Gratton}}}, \bibinfo {author} {\bibfnamefont
  {A.}~\bibnamefont {{Gregorio}}}, \bibinfo {author} {\bibfnamefont
  {A.}~\bibnamefont {{Gruppuso}}}, \bibinfo {author} {\bibfnamefont {J.~E.}\
  \bibnamefont {{Gudmundsson}}}, \bibinfo {author} {\bibfnamefont
  {J.}~\bibnamefont {{Hamann}}}, \bibinfo {author} {\bibfnamefont {F.~K.}\
  \bibnamefont {{Hansen}}}, \bibinfo {author} {\bibfnamefont {D.}~\bibnamefont
  {{Hanson}}}, \bibinfo {author} {\bibfnamefont {D.~L.}\ \bibnamefont
  {{Harrison}}}, \bibinfo {author} {\bibfnamefont {G.}~\bibnamefont {{Helou}}},
  \bibinfo {author} {\bibfnamefont {S.}~\bibnamefont {{Henrot-Versill{\'e}}}},
  \bibinfo {author} {\bibfnamefont {C.}~\bibnamefont
  {{Hern{\'a}ndez-Monteagudo}}}, \bibinfo {author} {\bibfnamefont
  {D.}~\bibnamefont {{Herranz}}}, \bibinfo {author} {\bibfnamefont {S.~R.}\
  \bibnamefont {{Hildebrand t}}}, \bibinfo {author} {\bibfnamefont
  {E.}~\bibnamefont {{Hivon}}}, \bibinfo {author} {\bibfnamefont
  {M.}~\bibnamefont {{Hobson}}}, \bibinfo {author} {\bibfnamefont {W.~A.}\
  \bibnamefont {{Holmes}}}, \bibinfo {author} {\bibfnamefont {A.}~\bibnamefont
  {{Hornstrup}}}, \bibinfo {author} {\bibfnamefont {W.}~\bibnamefont
  {{Hovest}}}, \bibinfo {author} {\bibfnamefont {Z.}~\bibnamefont {{Huang}}},
  \bibinfo {author} {\bibfnamefont {K.~M.}\ \bibnamefont {{Huffenberger}}},
  \bibinfo {author} {\bibfnamefont {G.}~\bibnamefont {{Hurier}}}, \bibinfo
  {author} {\bibfnamefont {A.~H.}\ \bibnamefont {{Jaffe}}}, \bibinfo {author}
  {\bibfnamefont {T.~R.}\ \bibnamefont {{Jaffe}}}, \bibinfo {author}
  {\bibfnamefont {W.~C.}\ \bibnamefont {{Jones}}}, \bibinfo {author}
  {\bibfnamefont {M.}~\bibnamefont {{Juvela}}}, \bibinfo {author}
  {\bibfnamefont {E.}~\bibnamefont {{Keih{\"a}nen}}}, \bibinfo {author}
  {\bibfnamefont {R.}~\bibnamefont {{Keskitalo}}}, \bibinfo {author}
  {\bibfnamefont {T.~S.}\ \bibnamefont {{Kisner}}}, \bibinfo {author}
  {\bibfnamefont {R.}~\bibnamefont {{Kneissl}}}, \bibinfo {author}
  {\bibfnamefont {J.}~\bibnamefont {{Knoche}}}, \bibinfo {author}
  {\bibfnamefont {L.}~\bibnamefont {{Knox}}}, \bibinfo {author} {\bibfnamefont
  {M.}~\bibnamefont {{Kunz}}}, \bibinfo {author} {\bibfnamefont
  {H.}~\bibnamefont {{Kurki-Suonio}}}, \bibinfo {author} {\bibfnamefont
  {G.}~\bibnamefont {{Lagache}}}, \bibinfo {author} {\bibfnamefont
  {A.}~\bibnamefont {{L{\"a}hteenm{\"a}ki}}}, \bibinfo {author} {\bibfnamefont
  {J.~M.}\ \bibnamefont {{Lamarre}}}, \bibinfo {author} {\bibfnamefont
  {A.}~\bibnamefont {{Lasenby}}}, \bibinfo {author} {\bibfnamefont
  {M.}~\bibnamefont {{Lattanzi}}}, \bibinfo {author} {\bibfnamefont {C.~R.}\
  \bibnamefont {{Lawrence}}}, \bibinfo {author} {\bibfnamefont {J.~P.}\
  \bibnamefont {{Leahy}}}, \bibinfo {author} {\bibfnamefont {R.}~\bibnamefont
  {{Leonardi}}}, \bibinfo {author} {\bibfnamefont {J.}~\bibnamefont
  {{Lesgourgues}}}, \bibinfo {author} {\bibfnamefont {F.}~\bibnamefont
  {{Levrier}}}, \bibinfo {author} {\bibfnamefont {A.}~\bibnamefont {{Lewis}}},
  \bibinfo {author} {\bibfnamefont {M.}~\bibnamefont {{Liguori}}}, \bibinfo
  {author} {\bibfnamefont {P.~B.}\ \bibnamefont {{Lilje}}}, \bibinfo {author}
  {\bibfnamefont {M.}~\bibnamefont {{Linden-V{\o}rnle}}}, \bibinfo {author}
  {\bibfnamefont {M.}~\bibnamefont {{L{\'o}pez-Caniego}}}, \bibinfo {author}
  {\bibfnamefont {P.~M.}\ \bibnamefont {{Lubin}}}, \bibinfo {author}
  {\bibfnamefont {J.~F.}\ \bibnamefont {{Mac{\'\i}as-P{\'e}rez}}}, \bibinfo
  {author} {\bibfnamefont {G.}~\bibnamefont {{Maggio}}}, \bibinfo {author}
  {\bibfnamefont {D.}~\bibnamefont {{Maino}}}, \bibinfo {author} {\bibfnamefont
  {N.}~\bibnamefont {{Mandolesi}}}, \bibinfo {author} {\bibfnamefont
  {A.}~\bibnamefont {{Mangilli}}}, \bibinfo {author} {\bibfnamefont
  {A.}~\bibnamefont {{Marchini}}}, \bibinfo {author} {\bibfnamefont
  {M.}~\bibnamefont {{Maris}}}, \bibinfo {author} {\bibfnamefont {P.~G.}\
  \bibnamefont {{Martin}}}, \bibinfo {author} {\bibfnamefont {M.}~\bibnamefont
  {{Martinelli}}}, \bibinfo {author} {\bibfnamefont {E.}~\bibnamefont
  {{Mart{\'\i}nez-Gonz{\'a}lez}}}, \bibinfo {author} {\bibfnamefont
  {S.}~\bibnamefont {{Masi}}}, \bibinfo {author} {\bibfnamefont
  {S.}~\bibnamefont {{Matarrese}}}, \bibinfo {author} {\bibfnamefont
  {P.}~\bibnamefont {{McGehee}}}, \bibinfo {author} {\bibfnamefont {P.~R.}\
  \bibnamefont {{Meinhold}}}, \bibinfo {author} {\bibfnamefont
  {A.}~\bibnamefont {{Melchiorri}}}, \bibinfo {author} {\bibfnamefont {J.~B.}\
  \bibnamefont {{Melin}}}, \bibinfo {author} {\bibfnamefont {L.}~\bibnamefont
  {{Mendes}}}, \bibinfo {author} {\bibfnamefont {A.}~\bibnamefont
  {{Mennella}}}, \bibinfo {author} {\bibfnamefont {M.}~\bibnamefont
  {{Migliaccio}}}, \bibinfo {author} {\bibfnamefont {M.}~\bibnamefont
  {{Millea}}}, \bibinfo {author} {\bibfnamefont {S.}~\bibnamefont {{Mitra}}},
  \bibinfo {author} {\bibfnamefont {M.~A.}\ \bibnamefont
  {{Miville-Desch{\^e}nes}}}, \bibinfo {author} {\bibfnamefont
  {A.}~\bibnamefont {{Moneti}}}, \bibinfo {author} {\bibfnamefont
  {L.}~\bibnamefont {{Montier}}}, \bibinfo {author} {\bibfnamefont
  {G.}~\bibnamefont {{Morgante}}}, \bibinfo {author} {\bibfnamefont
  {D.}~\bibnamefont {{Mortlock}}}, \bibinfo {author} {\bibfnamefont
  {A.}~\bibnamefont {{Moss}}}, \bibinfo {author} {\bibfnamefont
  {D.}~\bibnamefont {{Munshi}}}, \bibinfo {author} {\bibfnamefont {J.~A.}\
  \bibnamefont {{Murphy}}}, \bibinfo {author} {\bibfnamefont {P.}~\bibnamefont
  {{Naselsky}}}, \bibinfo {author} {\bibfnamefont {F.}~\bibnamefont {{Nati}}},
  \bibinfo {author} {\bibfnamefont {P.}~\bibnamefont {{Natoli}}}, \bibinfo
  {author} {\bibfnamefont {C.~B.}\ \bibnamefont {{Netterfield}}}, \bibinfo
  {author} {\bibfnamefont {H.~U.}\ \bibnamefont {{N{\o}rgaard-Nielsen}}},
  \bibinfo {author} {\bibfnamefont {F.}~\bibnamefont {{Noviello}}}, \bibinfo
  {author} {\bibfnamefont {D.}~\bibnamefont {{Novikov}}}, \bibinfo {author}
  {\bibfnamefont {I.}~\bibnamefont {{Novikov}}}, \bibinfo {author}
  {\bibfnamefont {C.~A.}\ \bibnamefont {{Oxborrow}}}, \bibinfo {author}
  {\bibfnamefont {F.}~\bibnamefont {{Paci}}}, \bibinfo {author} {\bibfnamefont
  {L.}~\bibnamefont {{Pagano}}}, \bibinfo {author} {\bibfnamefont
  {F.}~\bibnamefont {{Pajot}}}, \bibinfo {author} {\bibfnamefont
  {R.}~\bibnamefont {{Paladini}}}, \bibinfo {author} {\bibfnamefont
  {D.}~\bibnamefont {{Paoletti}}}, \bibinfo {author} {\bibfnamefont
  {B.}~\bibnamefont {{Partridge}}}, \bibinfo {author} {\bibfnamefont
  {F.}~\bibnamefont {{Pasian}}}, \bibinfo {author} {\bibfnamefont
  {G.}~\bibnamefont {{Patanchon}}}, \bibinfo {author} {\bibfnamefont {T.~J.}\
  \bibnamefont {{Pearson}}}, \bibinfo {author} {\bibfnamefont {O.}~\bibnamefont
  {{Perdereau}}}, \bibinfo {author} {\bibfnamefont {L.}~\bibnamefont
  {{Perotto}}}, \bibinfo {author} {\bibfnamefont {F.}~\bibnamefont
  {{Perrotta}}}, \bibinfo {author} {\bibfnamefont {V.}~\bibnamefont
  {{Pettorino}}}, \bibinfo {author} {\bibfnamefont {F.}~\bibnamefont
  {{Piacentini}}}, \bibinfo {author} {\bibfnamefont {M.}~\bibnamefont
  {{Piat}}}, \bibinfo {author} {\bibfnamefont {E.}~\bibnamefont {{Pierpaoli}}},
  \bibinfo {author} {\bibfnamefont {D.}~\bibnamefont {{Pietrobon}}}, \bibinfo
  {author} {\bibfnamefont {S.}~\bibnamefont {{Plaszczynski}}}, \bibinfo
  {author} {\bibfnamefont {E.}~\bibnamefont {{Pointecouteau}}}, \bibinfo
  {author} {\bibfnamefont {G.}~\bibnamefont {{Polenta}}}, \bibinfo {author}
  {\bibfnamefont {L.}~\bibnamefont {{Popa}}}, \bibinfo {author} {\bibfnamefont
  {G.~W.}\ \bibnamefont {{Pratt}}}, \bibinfo {author} {\bibfnamefont
  {G.}~\bibnamefont {{Pr{\'e}zeau}}}, \bibinfo {author} {\bibfnamefont
  {S.}~\bibnamefont {{Prunet}}}, \bibinfo {author} {\bibfnamefont {J.~L.}\
  \bibnamefont {{Puget}}}, \bibinfo {author} {\bibfnamefont {J.~P.}\
  \bibnamefont {{Rachen}}}, \bibinfo {author} {\bibfnamefont {W.~T.}\
  \bibnamefont {{Reach}}}, \bibinfo {author} {\bibfnamefont {R.}~\bibnamefont
  {{Rebolo}}}, \bibinfo {author} {\bibfnamefont {M.}~\bibnamefont
  {{Reinecke}}}, \bibinfo {author} {\bibfnamefont {M.}~\bibnamefont
  {{Remazeilles}}}, \bibinfo {author} {\bibfnamefont {C.}~\bibnamefont
  {{Renault}}}, \bibinfo {author} {\bibfnamefont {A.}~\bibnamefont {{Renzi}}},
  \bibinfo {author} {\bibfnamefont {I.}~\bibnamefont {{Ristorcelli}}}, \bibinfo
  {author} {\bibfnamefont {G.}~\bibnamefont {{Rocha}}}, \bibinfo {author}
  {\bibfnamefont {C.}~\bibnamefont {{Rosset}}}, \bibinfo {author}
  {\bibfnamefont {M.}~\bibnamefont {{Rossetti}}}, \bibinfo {author}
  {\bibfnamefont {G.}~\bibnamefont {{Roudier}}}, \bibinfo {author}
  {\bibfnamefont {B.}~\bibnamefont {{Rouill{\'e} d'Orfeuil}}}, \bibinfo
  {author} {\bibfnamefont {M.}~\bibnamefont {{Rowan-Robinson}}}, \bibinfo
  {author} {\bibfnamefont {J.~A.}\ \bibnamefont {{Rubi{\~n}o-Mart{\'\i}n}}},
  \bibinfo {author} {\bibfnamefont {B.}~\bibnamefont {{Rusholme}}}, \bibinfo
  {author} {\bibfnamefont {N.}~\bibnamefont {{Said}}}, \bibinfo {author}
  {\bibfnamefont {V.}~\bibnamefont {{Salvatelli}}}, \bibinfo {author}
  {\bibfnamefont {L.}~\bibnamefont {{Salvati}}}, \bibinfo {author}
  {\bibfnamefont {M.}~\bibnamefont {{Sandri}}}, \bibinfo {author}
  {\bibfnamefont {D.}~\bibnamefont {{Santos}}}, \bibinfo {author}
  {\bibfnamefont {M.}~\bibnamefont {{Savelainen}}}, \bibinfo {author}
  {\bibfnamefont {G.}~\bibnamefont {{Savini}}}, \bibinfo {author}
  {\bibfnamefont {D.}~\bibnamefont {{Scott}}}, \bibinfo {author} {\bibfnamefont
  {M.~D.}\ \bibnamefont {{Seiffert}}}, \bibinfo {author} {\bibfnamefont
  {P.}~\bibnamefont {{Serra}}}, \bibinfo {author} {\bibfnamefont {E.~P.~S.}\
  \bibnamefont {{Shellard}}}, \bibinfo {author} {\bibfnamefont {L.~D.}\
  \bibnamefont {{Spencer}}}, \bibinfo {author} {\bibfnamefont {M.}~\bibnamefont
  {{Spinelli}}}, \bibinfo {author} {\bibfnamefont {V.}~\bibnamefont
  {{Stolyarov}}}, \bibinfo {author} {\bibfnamefont {R.}~\bibnamefont
  {{Stompor}}}, \bibinfo {author} {\bibfnamefont {R.}~\bibnamefont
  {{Sudiwala}}}, \bibinfo {author} {\bibfnamefont {R.}~\bibnamefont
  {{Sunyaev}}}, \bibinfo {author} {\bibfnamefont {D.}~\bibnamefont {{Sutton}}},
  \bibinfo {author} {\bibfnamefont {A.~S.}\ \bibnamefont {{Suur-Uski}}},
  \bibinfo {author} {\bibfnamefont {J.~F.}\ \bibnamefont {{Sygnet}}}, \bibinfo
  {author} {\bibfnamefont {J.~A.}\ \bibnamefont {{Tauber}}}, \bibinfo {author}
  {\bibfnamefont {L.}~\bibnamefont {{Terenzi}}}, \bibinfo {author}
  {\bibfnamefont {L.}~\bibnamefont {{Toffolatti}}}, \bibinfo {author}
  {\bibfnamefont {M.}~\bibnamefont {{Tomasi}}}, \bibinfo {author}
  {\bibfnamefont {M.}~\bibnamefont {{Tristram}}}, \bibinfo {author}
  {\bibfnamefont {T.}~\bibnamefont {{Trombetti}}}, \bibinfo {author}
  {\bibfnamefont {M.}~\bibnamefont {{Tucci}}}, \bibinfo {author} {\bibfnamefont
  {J.}~\bibnamefont {{Tuovinen}}}, \bibinfo {author} {\bibfnamefont
  {M.}~\bibnamefont {{T{\"u}rler}}}, \bibinfo {author} {\bibfnamefont
  {G.}~\bibnamefont {{Umana}}}, \bibinfo {author} {\bibfnamefont
  {L.}~\bibnamefont {{Valenziano}}}, \bibinfo {author} {\bibfnamefont
  {J.}~\bibnamefont {{Valiviita}}}, \bibinfo {author} {\bibfnamefont
  {F.}~\bibnamefont {{Van Tent}}}, \bibinfo {author} {\bibfnamefont
  {P.}~\bibnamefont {{Vielva}}}, \bibinfo {author} {\bibfnamefont
  {F.}~\bibnamefont {{Villa}}}, \bibinfo {author} {\bibfnamefont {L.~A.}\
  \bibnamefont {{Wade}}}, \bibinfo {author} {\bibfnamefont {B.~D.}\
  \bibnamefont {{Wandelt}}}, \bibinfo {author} {\bibfnamefont {I.~K.}\
  \bibnamefont {{Wehus}}}, \bibinfo {author} {\bibfnamefont {M.}~\bibnamefont
  {{White}}}, \bibinfo {author} {\bibfnamefont {S.~D.~M.}\ \bibnamefont
  {{White}}}, \bibinfo {author} {\bibfnamefont {A.}~\bibnamefont
  {{Wilkinson}}}, \bibinfo {author} {\bibfnamefont {D.}~\bibnamefont {{Yvon}}},
  \bibinfo {author} {\bibfnamefont {A.}~\bibnamefont {{Zacchei}}}, \ and\
  \bibinfo {author} {\bibfnamefont {A.}~\bibnamefont {{Zonca}}},\ }\href
  {\doibase 10.1051/0004-6361/201525830} {\bibfield  {journal} {\bibinfo
  {journal} {\aap}\ }\textbf {\bibinfo {volume} {594}},\ \bibinfo {eid} {A13}
  (\bibinfo {year} {2016})},\ \Eprint {http://arxiv.org/abs/1502.01589}
  {arXiv:1502.01589 [astro-ph.CO]} \BibitemShut {NoStop}%
\bibitem [{\citenamefont {{van Belle}}\ \emph {et~al.}(2004)\citenamefont {{van
  Belle}}, \citenamefont {{Meinel}},\ and\ \citenamefont
  {{Meinel}}}]{vanBelle+2004_TelescopeCostScaling}%
  \BibitemOpen
  \bibfield  {author} {\bibinfo {author} {\bibfnamefont {G.~T.}\ \bibnamefont
  {{van Belle}}}, \bibinfo {author} {\bibfnamefont {A.~B.}\ \bibnamefont
  {{Meinel}}}, \ and\ \bibinfo {author} {\bibfnamefont {M.~P.}\ \bibnamefont
  {{Meinel}}},\ }in\ \href {\doibase 10.1117/12.552181} {\emph {\bibinfo
  {booktitle} {\procspie}}},\ \bibinfo {series} {Society of Photo-Optical
  Instrumentation Engineers (SPIE) Conference Series}, Vol.\ \bibinfo {volume}
  {5489},\ \bibinfo {editor} {edited by\ \bibinfo {editor} {\bibfnamefont
  {J.}~\bibnamefont {{Oschmann}}, \bibfnamefont {Jacobus~M.}}}\ (\bibinfo
  {year} {2004})\ pp.\ \bibinfo {pages} {563--570}\BibitemShut {NoStop}%
\bibitem [{\citenamefont {{Martinez}}\ \emph {et~al.}(2010)\citenamefont
  {{Martinez}}, \citenamefont {{Kolb}}, \citenamefont {{Sarazin}},\ and\
  \citenamefont {{Tokovinin}}}]{Martinze+2010_LargeTelescopeSeeing}%
  \BibitemOpen
  \bibfield  {author} {\bibinfo {author} {\bibfnamefont {P.}~\bibnamefont
  {{Martinez}}}, \bibinfo {author} {\bibfnamefont {J.}~\bibnamefont {{Kolb}}},
  \bibinfo {author} {\bibfnamefont {M.}~\bibnamefont {{Sarazin}}}, \ and\
  \bibinfo {author} {\bibfnamefont {A.}~\bibnamefont {{Tokovinin}}},\
  }\href@noop {} {\bibfield  {journal} {\bibinfo  {journal} {The Messenger}\
  }\textbf {\bibinfo {volume} {141}},\ \bibinfo {pages} {5} (\bibinfo {year}
  {2010})}\BibitemShut {NoStop}%
\bibitem [{\citenamefont {{Ofek}}(2019)}]{Ofek2019_Astrometry_Code}%
  \BibitemOpen
  \bibfield  {author} {\bibinfo {author} {\bibfnamefont {E.~O.}\ \bibnamefont
  {{Ofek}}},\ }\href {\doibase 10.1088/1538-3873/ab04df} {\bibfield  {journal}
  {\bibinfo  {journal} {\pasp}\ }\textbf {\bibinfo {volume} {131}},\ \bibinfo
  {pages} {054504} (\bibinfo {year} {2019})},\ \Eprint
  {http://arxiv.org/abs/1903.02015} {arXiv:1903.02015 [astro-ph.IM]}
  \BibitemShut {NoStop}%
\bibitem [{\citenamefont {{Brown}}\ and\ \citenamefont
  {{Twiss}}(1956)}]{Brown+Twiss1956_HBT_CorrelationsBetweenPhotons}%
  \BibitemOpen
  \bibfield  {author} {\bibinfo {author} {\bibfnamefont {R.~H.}\ \bibnamefont
  {{Brown}}}\ and\ \bibinfo {author} {\bibfnamefont {R.~Q.}\ \bibnamefont
  {{Twiss}}},\ }\href {\doibase 10.1038/177027a0} {\bibfield  {journal}
  {\bibinfo  {journal} {\nat}\ }\textbf {\bibinfo {volume} {177}},\ \bibinfo
  {pages} {27} (\bibinfo {year} {1956})}\BibitemShut {NoStop}%
\bibitem [{\citenamefont {{Brown}}\ and\ \citenamefont
  {{Twiss}}(1957)}]{Brown+Twiss1957_HBT_Interferometer}%
  \BibitemOpen
  \bibfield  {author} {\bibinfo {author} {\bibfnamefont {R.~H.}\ \bibnamefont
  {{Brown}}}\ and\ \bibinfo {author} {\bibfnamefont {R.~Q.}\ \bibnamefont
  {{Twiss}}},\ }\href {\doibase 10.1098/rspa.1957.0177} {\bibfield  {journal}
  {\bibinfo  {journal} {Proceedings of the Royal Society of London Series A}\
  }\textbf {\bibinfo {volume} {242}},\ \bibinfo {pages} {300} (\bibinfo {year}
  {1957})}\BibitemShut {NoStop}%
\bibitem [{\citenamefont {{Brown}}\ and\ \citenamefont
  {{Twiss}}(1958)}]{Brown+Twiss1958_HBT_Interferometer_Astronomy}%
  \BibitemOpen
  \bibfield  {author} {\bibinfo {author} {\bibfnamefont {R.~H.}\ \bibnamefont
  {{Brown}}}\ and\ \bibinfo {author} {\bibfnamefont {R.~Q.}\ \bibnamefont
  {{Twiss}}},\ }\href {\doibase 10.1098/rspa.1958.0239} {\bibfield  {journal}
  {\bibinfo  {journal} {Proceedings of the Royal Society of London Series A}\
  }\textbf {\bibinfo {volume} {248}},\ \bibinfo {pages} {199} (\bibinfo {year}
  {1958})}\BibitemShut {NoStop}%
\bibitem [{\citenamefont {{Eikenberry}}\ \emph {et~al.}(2019)\citenamefont
  {{Eikenberry}}, \citenamefont {{Bentz}}, \citenamefont {{Gonzalez}},
  \citenamefont {{Harrington}}, \citenamefont {{Jeram}}, \citenamefont {{Law}},
  \citenamefont {{Maccarone}}, \citenamefont {{Quimby}},\ and\ \citenamefont
  {{Townsend}}}]{Eikenberry+2019_PolyOculus}%
  \BibitemOpen
  \bibfield  {author} {\bibinfo {author} {\bibfnamefont {S.~S.}\ \bibnamefont
  {{Eikenberry}}}, \bibinfo {author} {\bibfnamefont {M.}~\bibnamefont
  {{Bentz}}}, \bibinfo {author} {\bibfnamefont {A.}~\bibnamefont {{Gonzalez}}},
  \bibinfo {author} {\bibfnamefont {J.}~\bibnamefont {{Harrington}}}, \bibinfo
  {author} {\bibfnamefont {S.}~\bibnamefont {{Jeram}}}, \bibinfo {author}
  {\bibfnamefont {N.}~\bibnamefont {{Law}}}, \bibinfo {author} {\bibfnamefont
  {T.}~\bibnamefont {{Maccarone}}}, \bibinfo {author} {\bibfnamefont
  {R.}~\bibnamefont {{Quimby}}}, \ and\ \bibinfo {author} {\bibfnamefont
  {A.}~\bibnamefont {{Townsend}}},\ }\href@noop {} {\bibfield  {journal}
  {\bibinfo  {journal} {arXiv e-prints}\ ,\ \bibinfo {eid} {arXiv:1907.08273}}
  (\bibinfo {year} {2019})},\ \Eprint {http://arxiv.org/abs/1907.08273}
  {arXiv:1907.08273 [astro-ph.IM]} \BibitemShut {NoStop}%
\bibitem [{\citenamefont {{Zackay}}\ and\ \citenamefont
  {{Gal-Yam}}(2014)}]{Zackay+2014_MultiplexerAlgo}%
  \BibitemOpen
  \bibfield  {author} {\bibinfo {author} {\bibfnamefont {B.}~\bibnamefont
  {{Zackay}}}\ and\ \bibinfo {author} {\bibfnamefont {A.}~\bibnamefont
  {{Gal-Yam}}},\ }\href {\doibase 10.1086/675390} {\bibfield  {journal}
  {\bibinfo  {journal} {\pasp}\ }\textbf {\bibinfo {volume} {126}},\ \bibinfo
  {pages} {148} (\bibinfo {year} {2014})},\ \Eprint
  {http://arxiv.org/abs/1310.3714} {arXiv:1310.3714 [astro-ph.IM]} \BibitemShut
  {NoStop}%
\bibitem [{\citenamefont {{Ben-Ami}}\ \emph {et~al.}(2014)\citenamefont
  {{Ben-Ami}}, \citenamefont {{Zackay}}, \citenamefont {{Rubin}}, \citenamefont
  {{Sagiv}}, \citenamefont {{Gal-Yam}},\ and\ \citenamefont
  {{Ofek}}}]{Ben-Ami+2014_MultiplexerInstrument}%
  \BibitemOpen
  \bibfield  {author} {\bibinfo {author} {\bibfnamefont {S.}~\bibnamefont
  {{Ben-Ami}}}, \bibinfo {author} {\bibfnamefont {B.}~\bibnamefont {{Zackay}}},
  \bibinfo {author} {\bibfnamefont {A.}~\bibnamefont {{Rubin}}}, \bibinfo
  {author} {\bibfnamefont {I.}~\bibnamefont {{Sagiv}}}, \bibinfo {author}
  {\bibfnamefont {A.}~\bibnamefont {{Gal-Yam}}}, \ and\ \bibinfo {author}
  {\bibfnamefont {E.~O.}\ \bibnamefont {{Ofek}}},\ }in\ \href {\doibase
  10.1117/12.2055421} {\emph {\bibinfo {booktitle} {\procspie}}},\ \bibinfo
  {series} {Society of Photo-Optical Instrumentation Engineers (SPIE)
  Conference Series}, Vol.\ \bibinfo {volume} {9147}\ (\bibinfo {year} {2014})\
  p.\ \bibinfo {pages} {91475U}\BibitemShut {NoStop}%
\end{thebibliography}%

\end{document}